\documentclass[12pt,preprint]{aastex}

\slugcomment{}

\shorttitle{A Study of the two Lynx Clusters at $z\sim1.3$}
\shortauthors{Jee et al.}

\begin{document}

\title{WEAK-LENSING DETECTION AT $z\sim1.3$:
MEASUREMENT OF THE TWO LYNX CLUSTERS WITH ADVANCED CAMERA FOR SURVEYS} 

\author{M.J. JEE\altaffilmark{1}, R.L. WHITE\altaffilmark{2}, H.C. FORD\altaffilmark{1},
        G.D. ILLINGWORTH\altaffilmark{3}
        J.P. BLAKESLEE\altaffilmark{4}, B. HOLDEN\altaffilmark{3}, AND S. MEI\altaffilmark{1}}

\altaffiltext{1}{Department of Physics and Astronomy, Johns Hopkins University, 3400 North Charles Street, Baltimore, MD 21218.}
\altaffiltext{2}{Space Telescope Science Institute, 3700 San Martin Drive, Baltimore, MD 21218.}
\altaffiltext{3}{University of California Observatories/Lick Observatory, University of California, Santa Cruz, CA 95064.}
\altaffiltext{4}{Department of Physics and Astronomy, Washington State University, WA 99164.}

\begin{abstract}
We present a {\it Hubble Space Telescope} Advanced Camera for Surveys (ACS) weak-lensing study of RX J0849+4452 and
RX J0848+4453, the two most distant (at $z=1.26$ and $z=1.27$, respectively)
clusters yet measured with weak-lensing. The two clusters
are separated by $\sim4\arcmin$ from each other and appear to form
a supercluster in the Lynx field.
Using our deep ACS $i_{775}$ and $z_{850}$
imaging, we detected weak-lensing signals around both clusters at
$\sim4\sigma$ levels.
The mass distribution indicated
by the reconstruction map is in good spatial agreement with the cluster galaxies.
From the singular isothermal sphere (SIS) fitting, we determined that RX J0849+4452 and RX J0848+4453 
have similar projected masses
of $(2.0\pm0.6)\times10^{14} M_{\sun}$ and $(2.1\pm0.7)\times10^{14} M_{\sun}$, respectively,
within a 0.5 Mpc ($\sim60\arcsec$) aperture radius. In order to
compare the weak-lensing measurements with the X-ray results calibrated by the
most recent low-energy quantum efficiency determination and time-dependent gain correction, 
we also re-analyzed the
archival $Chandra$ data and obtained $T=3.8_{-0.7}^{+1.3}$ and 
$1.7_{-0.4}^{+0.7}$~keV for RX J0849+4452 and RX J0848+4453, respectively.
Combined with the X-ray surface brightness profile measurement under the assumption
of isothermal $\beta$ model, the temperature of RX J0849+4452 predicts that the projected mass 
of the cluster within $r=0.5~$Mpc is $2.3_{-0.4}^{+0.8}\times10^{14} M_{\sun}$, consistent with the weak-lensing analysis.
On the other hand, for RX J0848+4453 we find that the mass derived from this X-ray analysis is
much smaller ($6.3_{-1.5}^{+2.6}\times10^{13} M_{\sun}$) than the weak-lensing measurement.
One possibility for this observed discrepancy is that the intracluster medium (ICM) 
of RX J0848+4453 has not yet fully thermalized. Although this interpretation is rather simplistic,
the relatively loose distribution of the cluster galaxies in part supports this possibility
of low degree of virialization. We also discuss other scenarios that might
give rise to the discrepancy.

\end{abstract}


\section{INTRODUCTION}

It has become clear that massive clusters are not extremely rare at high redshifts ($z>0.8$) and
the presence of these large collapsed structures when the age of the 
Universe is less than half its present value
is no longer in conflict with our current understanding
of the structure formation, especially in a
$\Lambda$-dominated flat cosmology.  
Pursuit of galaxy clusters to higher and higher redshift is important in
the extension of the evolutionary sequences to earlier epochs, when
the effect of the different cosmological frameworks becomes more discriminating.

A great deal of observational efforts have been made in the last decade
in enlarging the sample of high-redshift clusters. X-ray surveys have
provided an efficient method of cluster identification and probe of
cluster properties because a hot intracluster medium (ICM) within the cluster
generates strong diffuse X-ray emission and is believed to be in quasi-equilibrium
with gravity. However, it is questionable how well the clusters
selected by their X-ray excess can provide the unbiased representation of the typical
large scale structure at the cluster redshift. If 
the degree of the virialization decreases significantly with redshift and
is strongly correlated with X-ray temperature, the
cosmological dimming $\sim (1+z)^{-4}$ can bias our selection
progressively towards higher and higher mass, relaxed structures.

Among other important approaches to detect high-redshift clusters is a red-cluster-sequence (RCS)
survey using the distinctive spectral feature in cluster ellipticals. This
so-called 4000\AA~break feature is well-captured by
a careful combination of two passbands, and Gladders \& Yee (2005) recently reported
67 candidate clusters at a photometric redshift of $0.9 < z < 1.4$ from the $\sim10$\%  subregion
of the total $\sim100~\mbox{deg}^2$ RCS survey field. A related method
but giving a higher contrast of cluster members with respect to the background sources is
to use deep near-infrared (NIR) imaging (e.g., Stanford et al. 1997) for the selection of cluster 
candidates. High-redshift clusters identified in these color selection methods are
expected to serve as less biased samples encompassing the lower mass regime
at high redshifts.

In the current paper, we study two $z\sim1.3$ clusters, namely RX J0849+4452 
and RX J0848+4453 (hereafter Lynx-E and Lynx-W for brevity),
using the deep F775W and F850LP (hereafter
$i_{775}$ and $z_{850}$, respectively) images obtained with the Advanced Camera for Surveys (ACS)
on the $Hubble$ $Space$ $Telescope$ ($HST$). Interestingly, although these two clusters
are separated by only $\sim4\arcmin$ from each other, they were discovered by
different methods.

Standford et al. (1997) discovered Lynx-W in a NIR survey as an overdense
region of the $J-K > 1.9$ galaxies and spectroscopically confirmed 8 cluster members.
They also analyzed the archival ROSAT-PSPC observation of the region and found 
diffuse X-ray emission near the confirmed cluster galaxies. However, they
could not rule out the possibility that the X-ray flux might be coming
from the foreground point sources because of the PSPC PSF is too broad
to identify such objects. The subsequent study of the field using the $Chandra$ observations
showed that, although the previous ROSAT-PSPC observation is severely contaminated
by the X-ray point sources adjacent to the cluster, the cluster
is still responsible for some diffuse X-ray emission. Both
the X-ray temperature and luminosity 
of the cluster appear to be low ($T_X\sim1.6$~keV and $L_{bol}\sim0.69\times10^{44}
~ \mbox{ergs}~\mbox{s}^{-1}$; Stanford et al. 2001).

Lynx-E was, on the other hand, first discovered in the ROSAT Deep Cluster Survey (RDCS) as
a cluster candidate and follow-up near-infrared imaging showed an excess of
red ($1.8<J-K<2.1$) galaxies around the peak of the X-ray emission (Rosati et al. 1999). 
They also showed that five galaxies around the X-ray centroid have redshifts that are consistent
with the cluster redshift at $z=1.26$ using the Keck spectroscopic observations.
From the $Chandra$ data analysis, Stanford et al. (2001) determined the cluster temperature
and luminosity to be $T_X=5.8_{-1.7}^{+2.8}$ keV and $L_{bol}=3.3_{-0.5}^{+0.9}\times10^{44} \mbox{ergs}~\mbox{s}^{-1}$,
respectively.

The rather large difference in the X-ray properties of these two clusters may be viewed as
representing the characteristics of the sample obtained from
different survey methods. Lynx-E, the X-ray selected cluster, has
much higher X-ray temperature and luminosity than Lynx-W, the
NIR-selected cluster. If the stronger X-ray emission
means higher dynamical maturity, the more compact distribution
of the Lynx-E galaxies provides an alternate support of this
hypothesis. For dynamically relaxed systems the observed X-ray
properties can be easily translated into the mass properties
under the assumption of hydrostatic equilibrium. However,
as we probe into the higher and higher-redshift regime, it is natural
to expect that there will be more frequent occasions when
the equilibrium assumption loses its validity in deriving the
mass properties of the system. In addition, at $z>1$ we expect to
have a growing list of low-mass clusters that are also X-ray dark because
of the evasively low-temperature, as well as the substantial cosmological dimming.
Therefore, it is plausible to suspect 
that these two $z\sim1.3$ clusters (especially Lynx-W, the poorer X-ray system)
might lie on a border where the X-ray observations alone start
to become insufficient to infer the mass properties.

Weak-lensing provides an alternative approach to
deriving the mass of a gravitationally bound
system without relying on assumptions about the dynamical state. This
can help us to probe the properties of the high-redshift
clusters in lower mass regimes, where the X-ray measurements
alone may not provide useful physical quantities.
In our particular case, weak-lensing is an important tool to test how
the masses of the two Lynx clusters at $z\sim1.3$
compare with their X-ray measurements. Especially
for Lynx-W, weak-lensing seems to be the unique route for probing
the cluster mass, considering the poor and amorphous X-ray emission.
Another interesting question is whether the low X-ray temperature of Lynx-W arises simply from a low
mass or from a yet poor thermalization of the ICM.

However, the detection of weak-lensing signal at $z\sim1.3$ 
is difficult and much more so if the lens is not very massive.
In our previous investigation of the two $z\sim0.83$ high-redshift
clusters (Jee et al. 2005a, hereafter Paper I; Jee et al. 2005b, hereafter Paper II),
we were able to detect clear lensing signals. They revealed the
complicated dark matter substructure of the clusters in great detail.
The effective source plane (defined by the effective mean redshift of the background galaxies)
in Paper I and II is at $z_{eff}\sim1.3$, corresponding to
the redshift of the lenses targeted in the current paper! Therefore, the
number density of background galaxies decreases substantially
compared to our $z\sim0.8$ studies and, in addition, the higher fraction of
non-background population in our source sample inevitably dilutes the
resulting lensing signal quite severely. Furthermore, the accurate
removal of instrumental artifacts becomes more critical as stronger
signals come from more distant, and thus fainter and smaller galaxies.
They are more severely affected by the point-spread-function (PSF).
Nevertheless, our
analyses of RDCS 1252-2927 at $z=1.24$ (Lombardi et al. 2005; Jee et al. in
preparation) demonstrate that weak-lensing can still be 
applied to clusters even at these redshifts and
reveals the cluster mass distribution with high significance.

Returning to the X-ray properties, the low-energy quantum efficiency (QE) degradation of the $Chandra$ instrument can
cause noticeable biases in cluster temperature
measurements. Although there have been many suggestions regarding this issue,
it was not until recently that a convergent prescription to remedy
the situation has become available from the $Chandra$ 
X-ray Center\footnote{see http://cxc.harvard.edu/ciao3.0/threads/apply\_acisabs/
or http://cxc.harvard.edu/ciao3.2/releasenotes/}.
Because we suspect that the previous X-ray analyses of the Lynx clusters
suffered from the relatively insufficient understanding of this problem, we have also re-analyzed
the archival $Chandra$ data to enable a fairer
comparison between the weak-lensing and X-ray measurements.

Throughout the paper, we assume a $\Lambda$CDM cosmology favored by the 
Wilkinson Microwave Anisotropy Probe (WMAP), where $\Omega_M$, $\Omega_{\Lambda}$, and
$H_0$ are 0.27, 0.73, and 71 $\mbox{km}~\mbox{s}^{-1}~\mbox{Mpc}^{-1}$, respectively.
All the quoted uncertainties are at the 1 $\sigma$ ($\sim68$\%) level.

\section{OBSERVATIONS}
\subsection{ACS Observation \label{subsection_acsobservation}}
Deep ACS/WFC imaging of the Lynx clusters were carried out as part of ACS Guaranteed Time Observation (GTO)
during 2004 March in three contiguous pointings, which cover a strip of $\sim9\arcmin\times3\arcmin$ region.
A slight overlap ($\sim30\arcsec$) was made between the pointings and the strip is oriented in
such a way that the two cluster centers are approximately located near the overlap region.
Each pointing was observed in $i_{775}$ and $z_{850}$
passbands with 3 and 5 orbits of integration, respectively.

We used the ACS GTO pipeline (``APSIS"; Blakeslee et al. 2003) to remove cosmic rays, correct
geometric distortion via drizzle algorithm (Fruchter and Hook 2002), and register different exposures. Apsis 
meets all the requirements of weak lensing analysis (Paper I and II), offering
a precise ($\sim0.015$ pixels) image registration via the ``match'' program (Richmond 2002)
after correcting for geometric distortion (Meurer et al. 2003).

\begin{figure}
\plotone{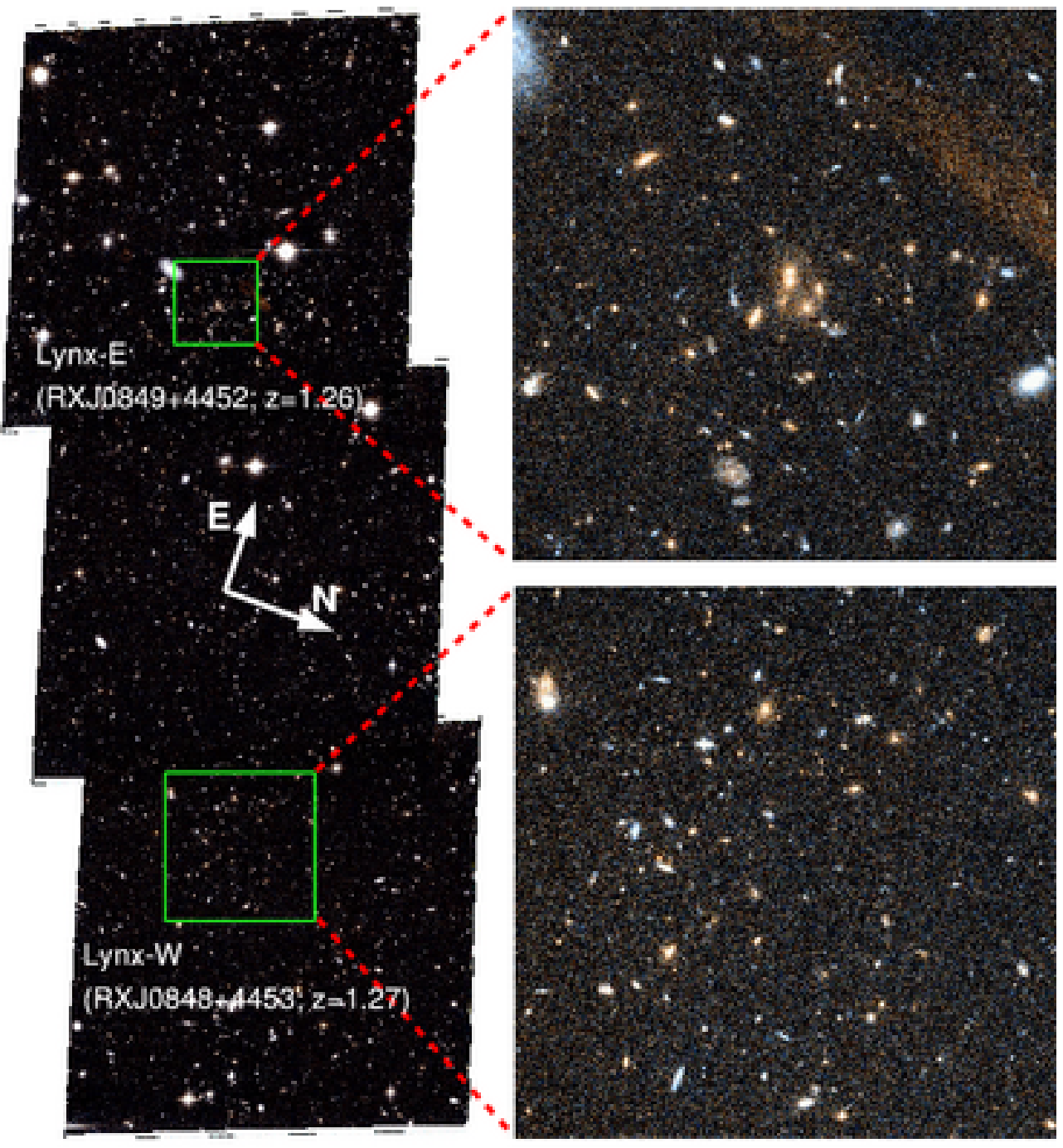}
\caption{Pseudo-color composite image created from the ACS $i_{775}$ and $z_{850}$
passband images. The approximate locations of the two high-redshift Lynx clusters
are illustrated by the green boxes and we present their blow-ups on the right.
The excess of the cluster red sequence is obvious in both clusters, though
Lynx-E has the dominant BCGs and the distribution of the galaxies seems to be more compact.
The orientation of the image shown by the compass is maintained throughout the paper
unless commented otherwise.
}
\label{fig_lynx_illustration}
\end{figure}

In Figure~\ref{fig_lynx_illustration} we present the pseudo-color 
image of the entire ACS field with the blow-ups of the two Lynx clusters.
Lynx-E is well-portrayed by the somewhat compact distribution
of the cluster red sequence around the brightest cluster galaxies (BCGs).
It appears that the cluster has a strongly lensed blue giant arc $\sim4.5\arcsec$
south of the BCGs. The spectroscopic
redshift of this arc candidate has not yet been determined. 
The red sequence of Lynx-W looks somewhat scattered and there
seem to be no distinct BCGs characterizing the cluster center though
the excess of the early-type galaxies in the region clearly defines
the cluster locus.

The detection image was created by combining the two passband images using inverse variance
weighting. Objects are detected through the SExtractor program (Bertin \& Arnouts 1996) by searching for 
at least five connected pixels above 1.5 times the sky rms. The field contains several
bright stars whose diffraction spikes not only induce a false detection, but
also contaminate the neighboring objects. We manually selected and removed these objects.
The catalog contains a total of 8737 galaxies.

\subsection{Chandra Observation}
We retrieved the $Chandra$ observation of the Lynx field from the
Chandra X-ray Center. The field was observed with the Advanced CCD Imaging
Spectrometer I-array (ACIS-I) in the faint mode at a focal temperature of -120 K.
The observation consists of two exposures: $\sim65$ ks and $\sim125$ ks integrations
on 2000 May 3 and 4, respectively.
The raw X-ray events were processed with the $Chandra$ Interactive Analysis of 
Observations (CIAO) software version 3.2 and the Calibration Database (CALDB) version 3.1, 
which provide the correction for time-dependent gain variation and the low-energy
quantum efficiency degradation without requiring any external guidance.
We identified and flagged hot pixels and afterglow events using the $acis\_build\_badpix$,
$acis\_classify\_hotpix$ and $acis\_find\_hotpix$ scripts while selecting only the standard 
$ASCA$ events (0,2,3,4, and 6).

\begin{figure}
\plotone{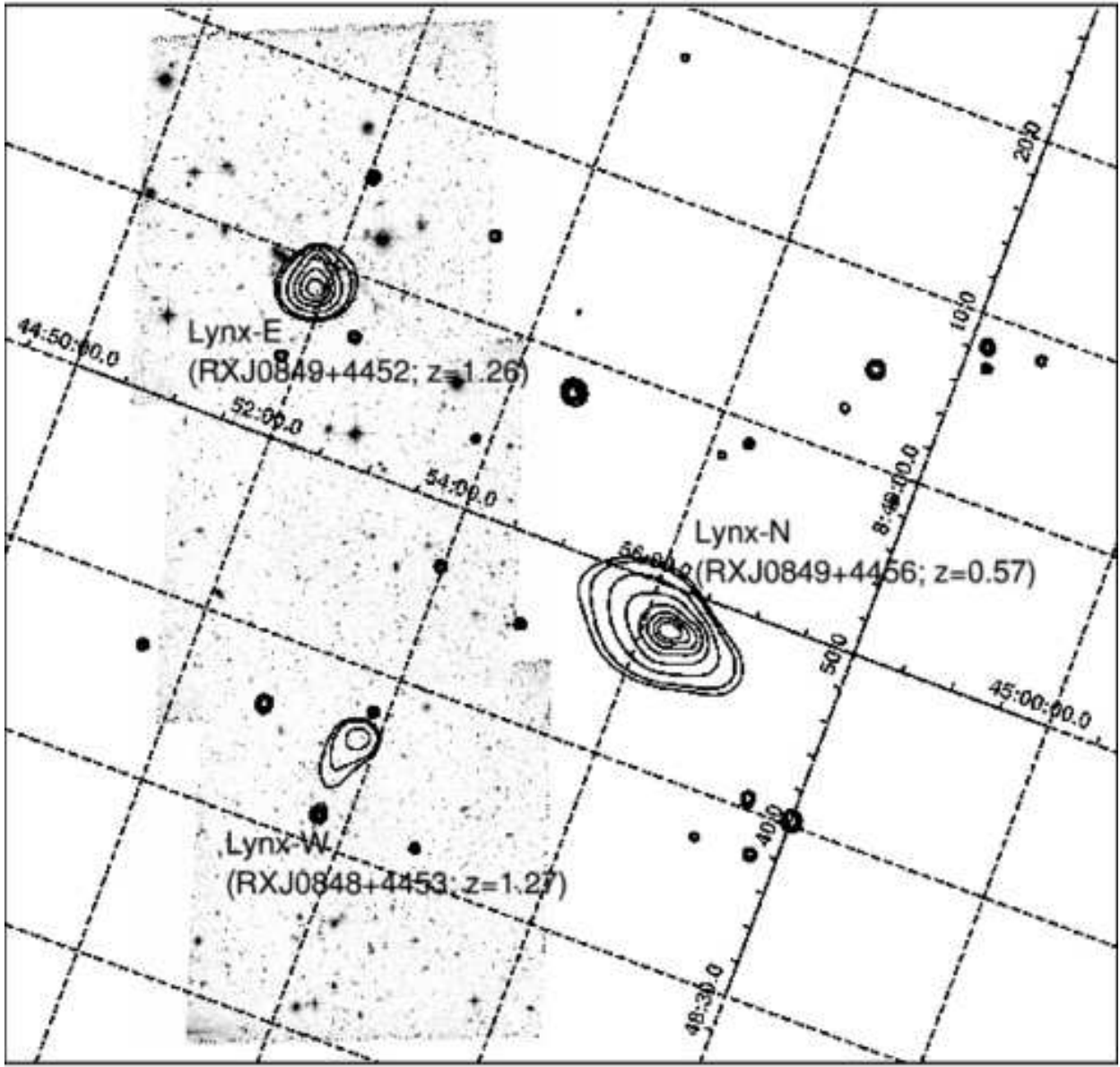}
\caption{Chandra X-ray contours overlaid on top of the ACS image. The X-ray
image is adaptively smoothed with the CIAO's CSMOOTH program in such a way
that the minium significance is 3$\sigma$. The contours are spaced
in square-root scale. The two high-redshift clusters are in
good spatial agreement with the cluster galaxies shown in Figure~\ref{fig_lynx_illustration}.
The foreground cluster (RXJ0849+4456) at $z=0.57$ is outside the ACS pointings.
}
\label{fig_xrayoverimage}
\end{figure}

Figure~\ref{fig_xrayoverimage} shows the adaptively smoothed $Chandra$ X-ray contours
of the Lynx field overlaid on the ACS image. This adaptive smoothing
is performed using the CIAO CSMOOTH program with a minimum significance of 3 $\sigma$ and
the contours are spaced in square-root scale. Because of the low counts from the two
high-redshift clusters, the 3 $\sigma$ significance condition can only be met
with rather large smoothing kernels. Therefore, the round appearance
of the contours should not be misinterpreted as indicating the relaxed status of the systems.
When the contours are reproduced with a smaller, constant kernel smoothing, Lynx-W
looks much more irregular than Lynx-E.
The X-ray centroids of Lynx-E and W are in good spatial
agreement with those of cluster optical lights.
The foreground cluster RXJ 0849+4456 (Holden et al. 2001) at z=0.57 
appears to be also strong in X-ray emission, but is located
outside the ACS pointings ($\sim5\arcmin$ and $\sim3\arcmin$ apart
from Lynx-E and W, respectively). The multi-wavelength analysis of this
cluster is presented by Holden et al. (2001) and they found that
the cluster can be further resolved into two groups at z=0.57 and 0.54.
Our subsequent X-ray analyses are confined to the two
high-redshift clusters at $\bar{z}=1.265$ present within the current ACS pointings.

\section{ACS DATA ANALYSIS}
As in our previous investigations (Paper I and II), we measure galaxy shapes
and model the point-spread-function of the observation using shapelets (Bernstein \& Jarvis 2002; Refregier 2003).
Readers are referred to Paper I and II for detailed description of the ellipticity measurements.

\subsection{Cluster Luminosity}
Our current spectroscopic catalog of the ACS Lynx field (B. Holden et al. in preparation) contains
150 objects and 32 of them belong to either of the two high-redshift clusters ($1.24<z<1.28$);12 galaxies 
are at $z>1.31$ and the rest of them (106 objects) are foreground objects.

We supplemented the cluster member galaxy catalog with
the cluster red sequence (Mei et al. 2005) using 
$i_{775}-z_{850}$ colors. In order to minimize the systematics 
from internal gradients and the different
PSF sizes (the PSF of $z_{850}$ is $\sim10$\% broader than that of $i_{775}$), the galaxies are deconvolved 
with the CLEAN (H\"{o}gbom et al. 1974) algorithm. After an effective radius $R_e$ is determined for each galaxy, we
measured the object colors within a circular aperture defined by $R_e$. When the estimated $R_e$ was less than
three pixels, we used a three pixel aperture instead (the median $R_e$ is $\sim5$~pixels).

\begin{figure}
\plotone{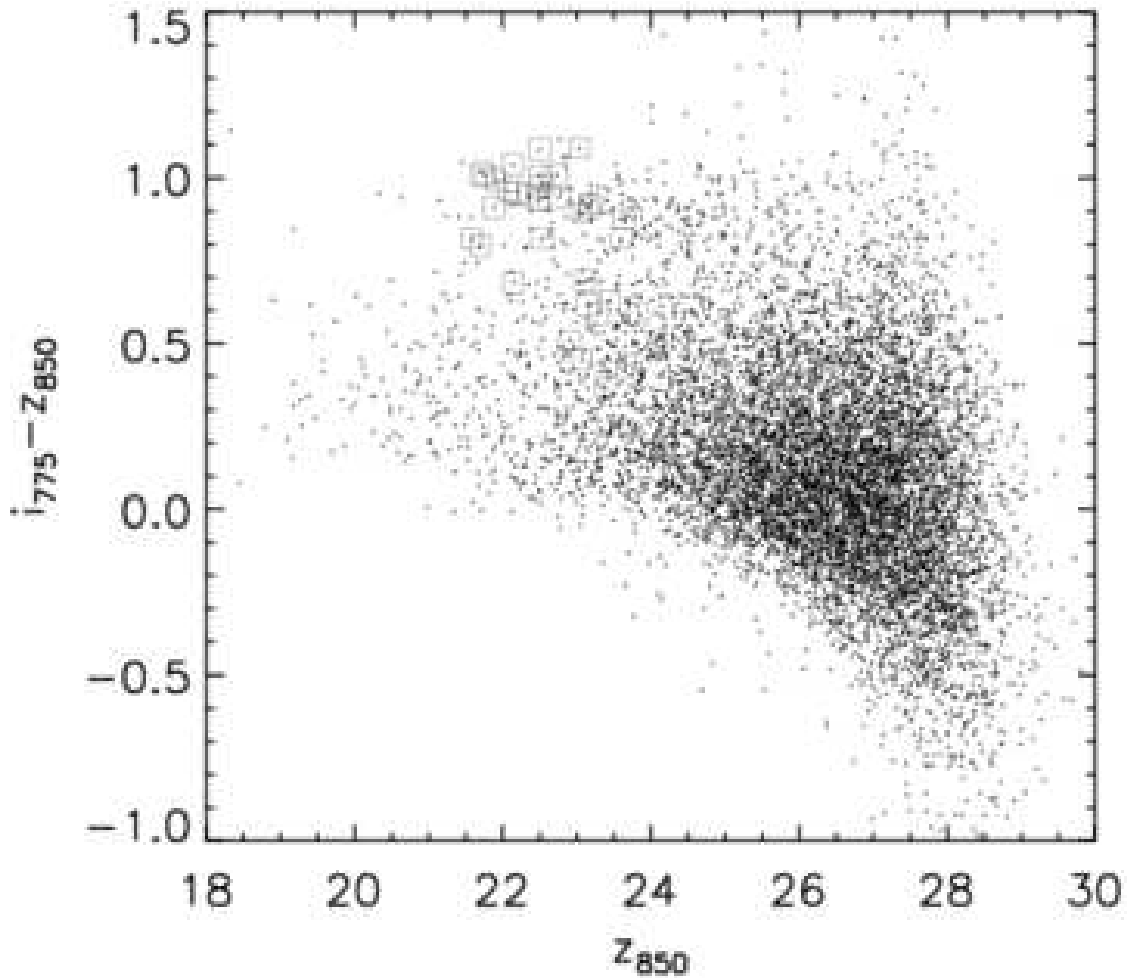}
\caption{The color-magnitude diagram of the ACS Lynx field. The $z_{850}$ band
filter is centered somewhat blueward of the rest-frame 4000 \AA~break
at z=1.265, making the $i_{775}-z_{850}$ color less than ideal for
determining the rest-frame U-B color. The red sequence, however, is 
still visible with the approriate scatter at $i_{775}-z_{850}\sim1.0$ (see Mei et al. 2005).
We place squares on the known cluster members ($1.25<z<1.28$).
}
\label{fig_cm}
\end{figure}

At $z\sim1.265$, the 4000\AA~break
is shifted slightly blue-ward of the effective wavelength of the $z_{850}$ filter. Therefore, 
this filter combination is less than ideal, but
the red sequence is still visible	
down to $z_{850}\sim24$ in the $i_{775}-z_{850}$ versus $z_{850}$ plot (Figure~\ref{fig_cm}).
We visually examined each candidate and discarded the objects
that do not seem to have early-type morphology, or whose
redshifts (if known) are inconsistent with the cluster redshifts.
The final cluster member catalog contains 68 objects.

The rest-frame $B$ band at the cluster redshift is approximately
redshifted to the ACS $z_{850}$ band and we derive the
following photometric transformation from the synthetic photometry
with the Spectral Energy Distribution (SED) templates of Kinney et al. (1996).
\begin{equation}
B_{rest} = z_{850} - (0.70 \pm 0.02) (i_{775}-z_{850}) + (1.08 \pm 0.01) - DM \label{photran},
\end{equation}
\noindent
where DM is the distance modulus of 44.75 at $\bar{z}=1.265$.

From the above selection of the cluster galaxies,
we estimate that Lynx-E and W encloses $L_B\sim1.5\times10^{12}$ and
$\sim0.8\times10^{12} L_{B\sun}$, respectively within 0.5 Mpc ($\sim60\arcsec$) radius. 
Of course, these values
correspond to the lower limits because we neglected the
contribution from the blue galaxies (except for the several spectroscopically confirmed
ones), as well
as the less luminous population ($z_{850}>24$).
However, we do not attempt to determine the correction factors in the current
paper because the number of galaxies in both of our spectroscopic
and red sequence samples is insufficient to 
support our statistical derivation.

\subsection{PSF Correction}
ACS/WFC has a time- and position-dependent PSF (Paper I) and the
ability to properly model the PSF pattern in the observed cluster field is critical
in subsequent galaxy ellipticity analysis. In paper I and II, we demonstrated that 
the PSF of WFC sampled from the 47 Tucanae field can be used to describe the
PSF pattern of the cluster images where only a limited number of stars are available, but can
be used as a diagnosis of the model accuracy.

\begin{figure}
\plotone{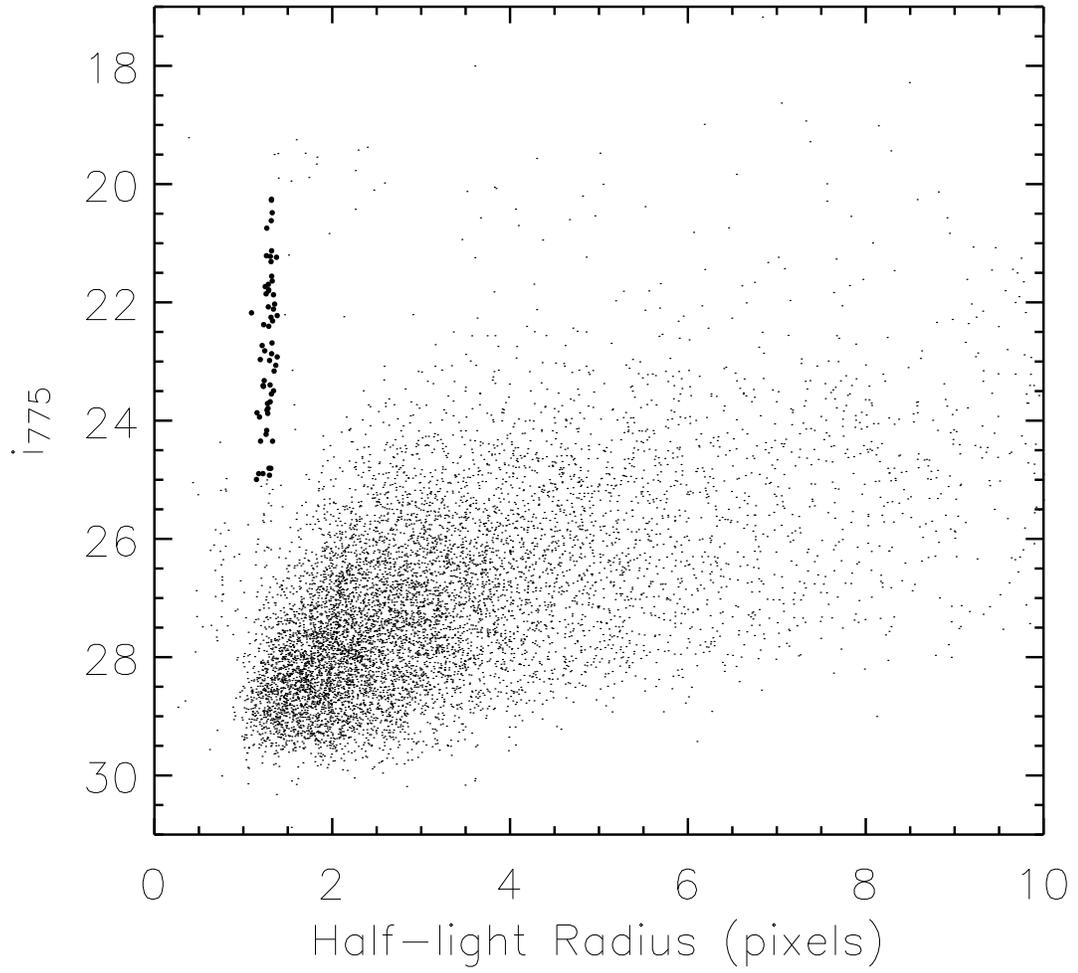}
\caption{The galaxy-star separation. The stars (large symbols) occupy a narrow locus in
this size (half-light radius) versus magnitude plot. 
}
\label{fig_starselect}
\end{figure}

\begin{figure}
\plottwo{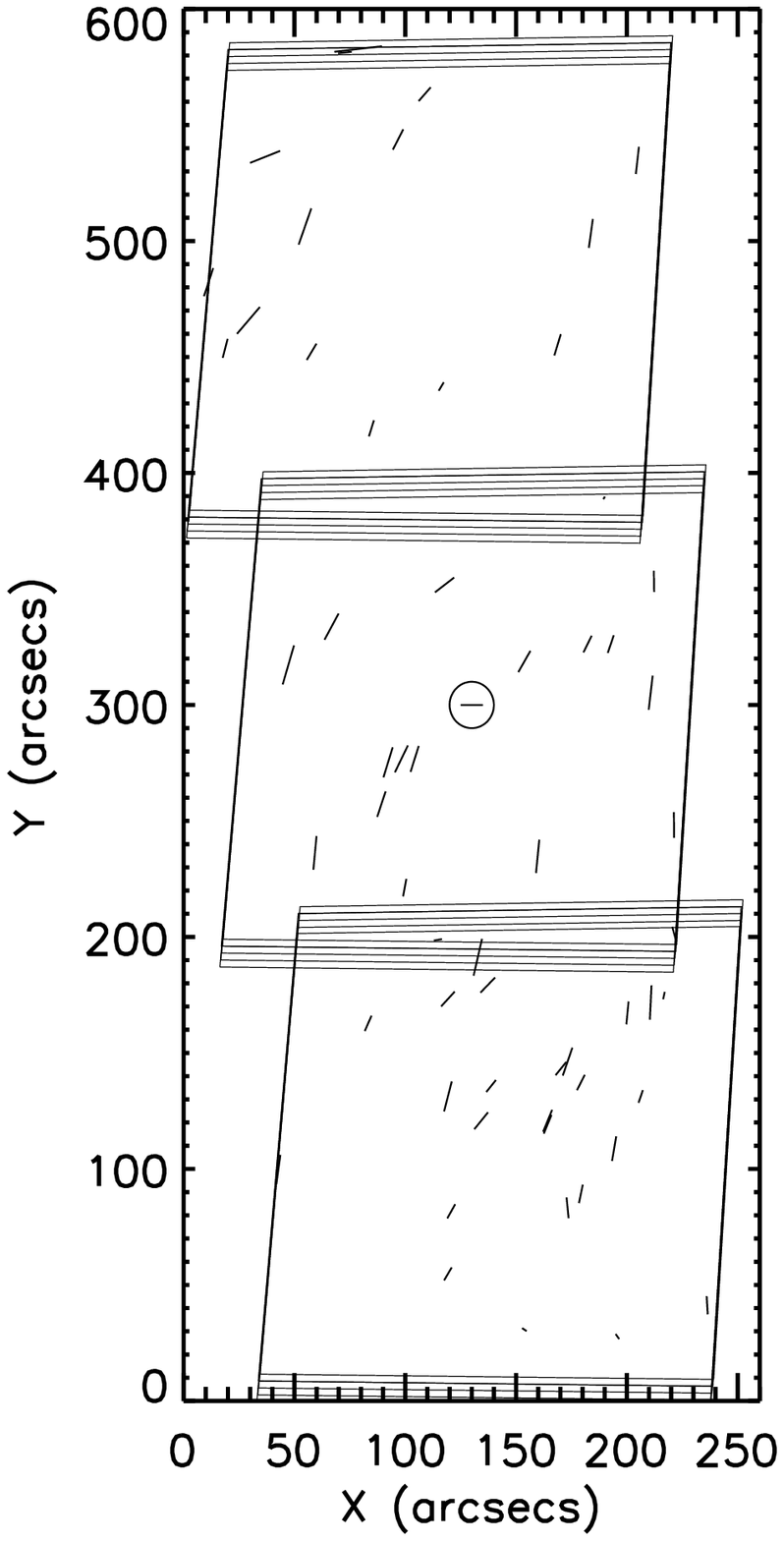}{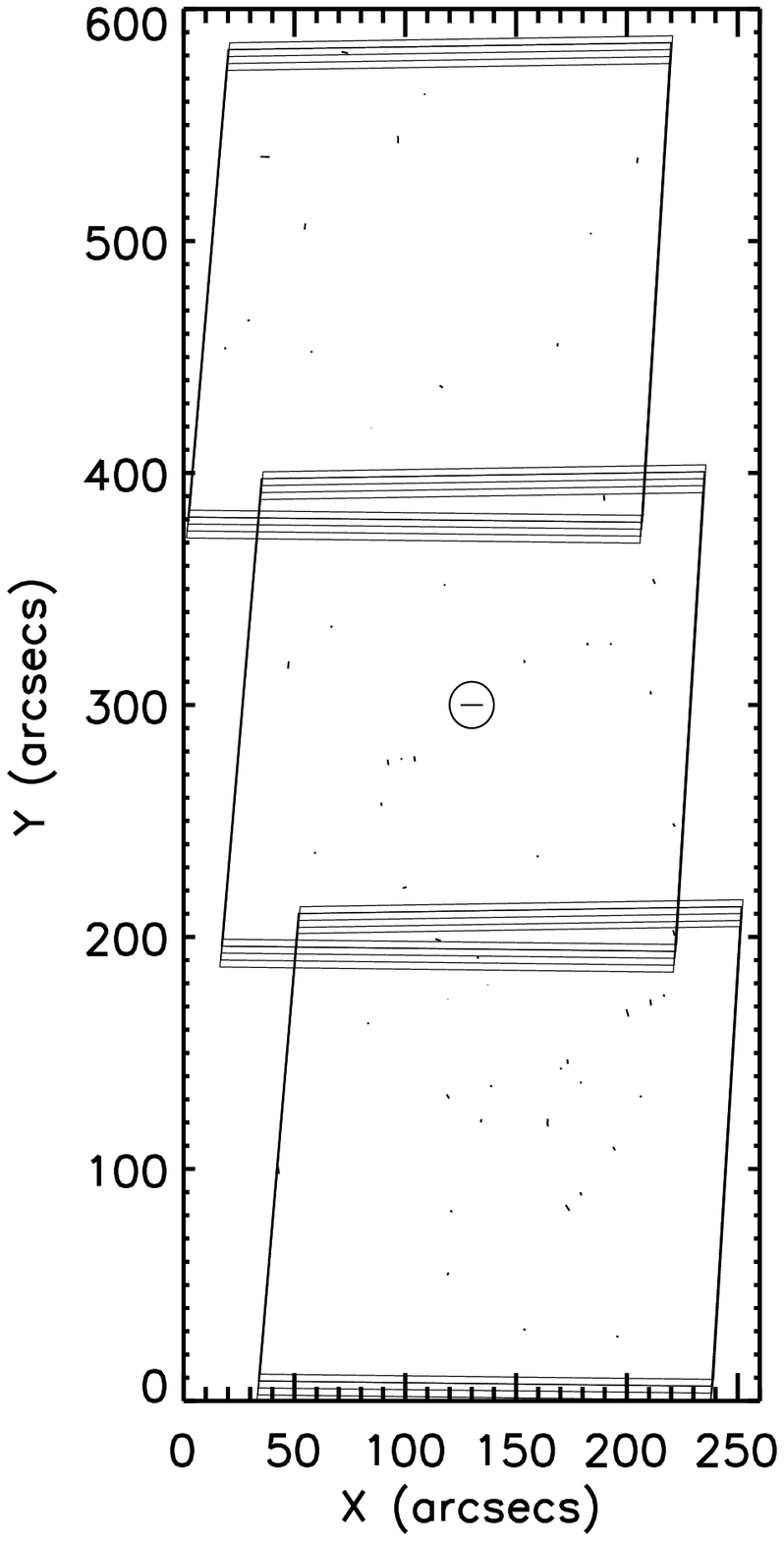}
\caption{Ellipticity of the stars present in the ACS Lynx field ($i_{775}$) before (a) and
after (b) the PSF correction. The stars show a typical ellipticity
of $\delta\simeq0.1$ (a). When the images are convolved with
a rounding kernel, the resulting residual ellipticity decreases to $\delta\simeq0.02$ (b).
The test verifies that our PSF variation model closely matches
the real PSF pattern in the ACS Lynx field.
}
\label{fig_starfield}
\end{figure}

We selected the stars in the Lynx field via a typical magnitude versus half-light radius plot 
(Figure~\ref{fig_starselect}).  Figure~\ref{fig_starfield}a show the WFC PSF pattern in the $i_{775}$ image
of the Lynx field, which is
similar to the ones in our previous cluster weak lensing studies. The PSFs are elongated
in the lower-left to upper-right direction. An analogous pattern is also observed
in the $z_{850}$ band. However, the wings of the $z_{850}$
are stretched approximately parallel to the row of the CCD (telescope V2 axis) and
the feature becomes observable when the wings of the PSFs are more heavily weighed (Heymans et al. 2005).

In our calibration of the ACS (Sirianni et al. 2005),
we also observed an opposite pattern (i.e., with an ellipticity nearly perpendicular to
Figure~\ref{fig_starfield}a), it seems that this PSF pattern is more frequently observed, at least
in our GTO surveys of $\sim15$ clusters. Because the focus offsets of different HST visits are
likely to vary, one may desire to find the closest PSF template 
for every individual exposure and perform PSF corrections one by one.
However, we find that in our GTO cluster observations the PSF patterns in different exposures do not 
vary considerably. Therefore, we chose a single PSF template for each filter and 
created a PSF map for the entire $3\times 1$ mosaic image by placing the template PSFs on each pointing.
In order to minimize the model-data discrepancy due to the slight focus variation, we fine-tuned our model
for each exposure by shearing
the PSF by an amount $\delta \eta$, which can be expressed in shapelet notation as
\begin{equation}
b_{pq}^{\prime} = \mbox{\bf{S}}_{\delta \eta} b_{pq},
\end{equation}
\noindent
where $b_{pq}$ is the shapelet componet of the PSF and the evaluation of matrix elements of the shear operator 
$\mbox{\bf{S}} _ {\delta\eta}$ can be found in Bernstein \& Jarvis (2002).

\begin{figure}
\plotone{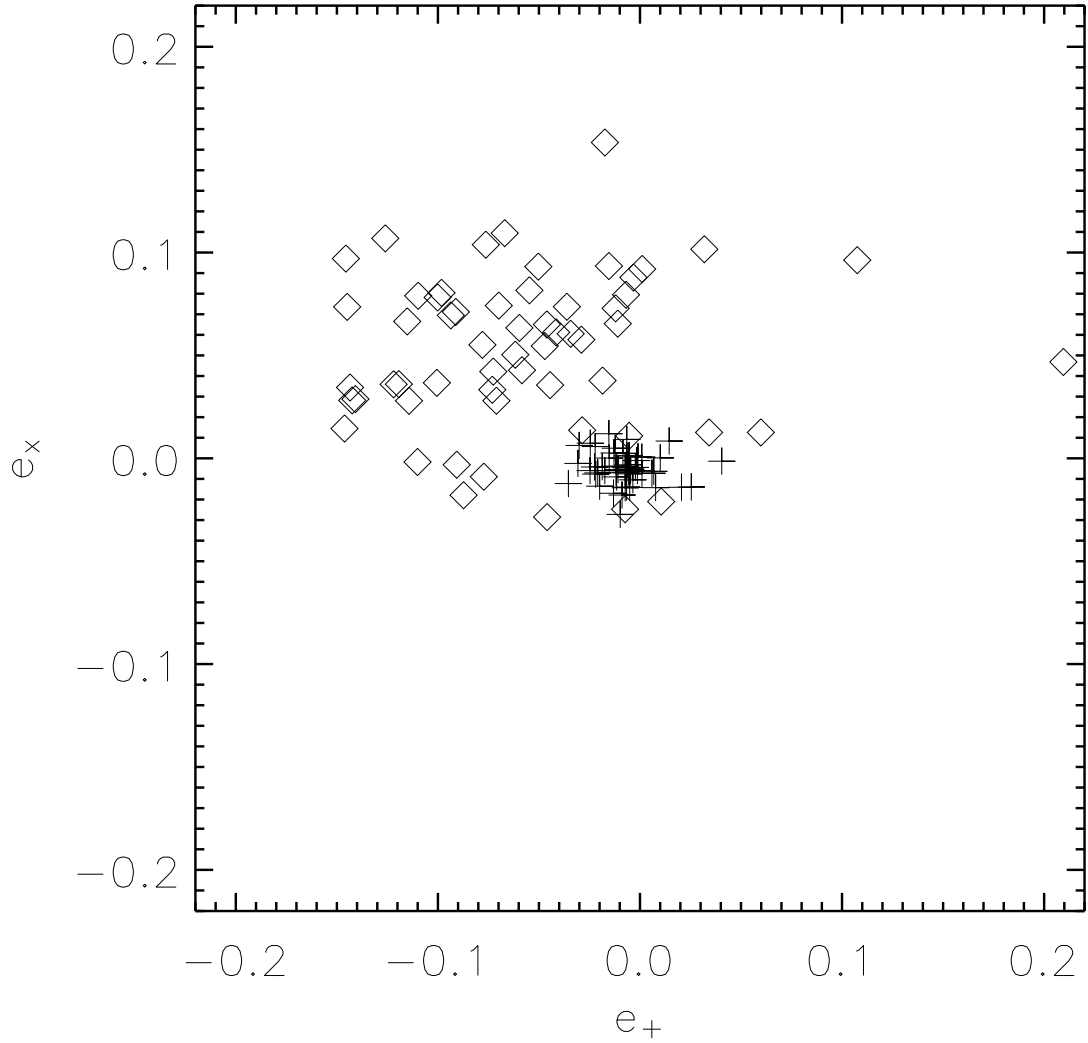}
\caption{Same as in Figure~\ref{fig_starfield}. The ellipcities are now
represented in their $e_+$ (cross) and $e_\times$ (diagonal) components.
The diamond and plus symbols show the the ellipticity
of the stars before and after the PSF correction, respectively.
\label{fig_star_anisotropy}}

\end{figure}

Figure~\ref{fig_starfield}b displays
the residual ellipticities of the same stars in the $i_{775}$ when the PSF is circularized 
with rounding kernels (Fischer \&
Tyson 1997; Kaiser 2000; Bernstein \& Jarvis 2002). The dramatic reduction
of the PSF anisotropy is also distinct when the ellipticity components ($e_+$ and $e_{\times}$) before
and after the corrections are compared (Figure~\ref{fig_star_anisotropy}).

This rounding kernel test verifies that our PSF models describe the PSF pattern of the
cluster observation very precisely. Although one can continue with this rounding kernel
method and make a subsequent measurement of the galaxy shape in this ``rounded'' images (e.g., Fischer \& Tyson 1997),
we prefer to remove the PSF effect through straightforward deconvolution in $shapelets$ because the latter
gives more satisfactory results for very faint galaxies (Paper I; Hirata \& Seljak 2003). Besides,
the $z_{850}$ PSF is rather complicated because of the ellipticity variation between core and wing metioned above,
and this PSF effect can be more efficiently corrected by the deconvolution.

\subsection{Mass Reconstruction \label{section_mass_reconstruction}}
In order to maximize the weak-lensing signal, it is important to select the source population
in such a way that the source sample contains the minimal contamination from cluster and foreground
galaxies. Because only two passband images of the Lynx field are available,
direct determination of reliable photometric redshift for an individual galaxy is impossible. 
Therefore,
we chose to select the background galaxies based on their ($i_{755}-z_{850}$) colors and $z_{850}$
magnitudes. The redshift distribution of this sample can be indirectly inferred when we
apply the same selection criteria to other deep multi-band HST observations such as the
Ultra Deep Field (UDF; Beckwidth et al. 2003) project, for which reliable photometric redshift information is obtainable down
to the limiting magnitude of our cluster observation (D. Coe et al., in preparation).

We selected the $24<z_{850}<28.5$ galaxies whose $i_{775}-z_{850}$ colors are
bluer than those of the cluster redsequence ($i_{775}-z_{850}\lesssim0.7)$ as ``optimal'' background population by examining
the resulting tangential shears around the two $\bar{z}=1.265$ clusters. 
This selection yields a total of 6742 galaxies ($\sim204 \mbox{arcmin}^{-2}$).
Assuming that the cosmic variance between the Lynx and UDF is not large,
we estimate that approximately 60 per cent of the selection is behind the Lynx clusters.

Our final ellipticity catalog was created by combining the $i_{775}$ and $z_{850}$ bandpass ellipticities.
Of course, there is a subtlety in this procedure because an object can have intrinsically
different shapes and thus ellipticities in different passbands. We adopted the methodology presented by
Bernstein and Jarvis (2002) to optimally combine the galaxy ellipticities.
In our previous weak-lensing analyses (Paper I and II), we found that this scheme indeed reduced the
mass reconstruction scatters compared to the case when only single passband images were used; the improvement
increases as fainter galaxies are included.
As a consistency check, we compared the shapes and lensing signals from the two passband images, and
confirmed that the results are statistically consistent. 

\begin{figure}
\plottwo{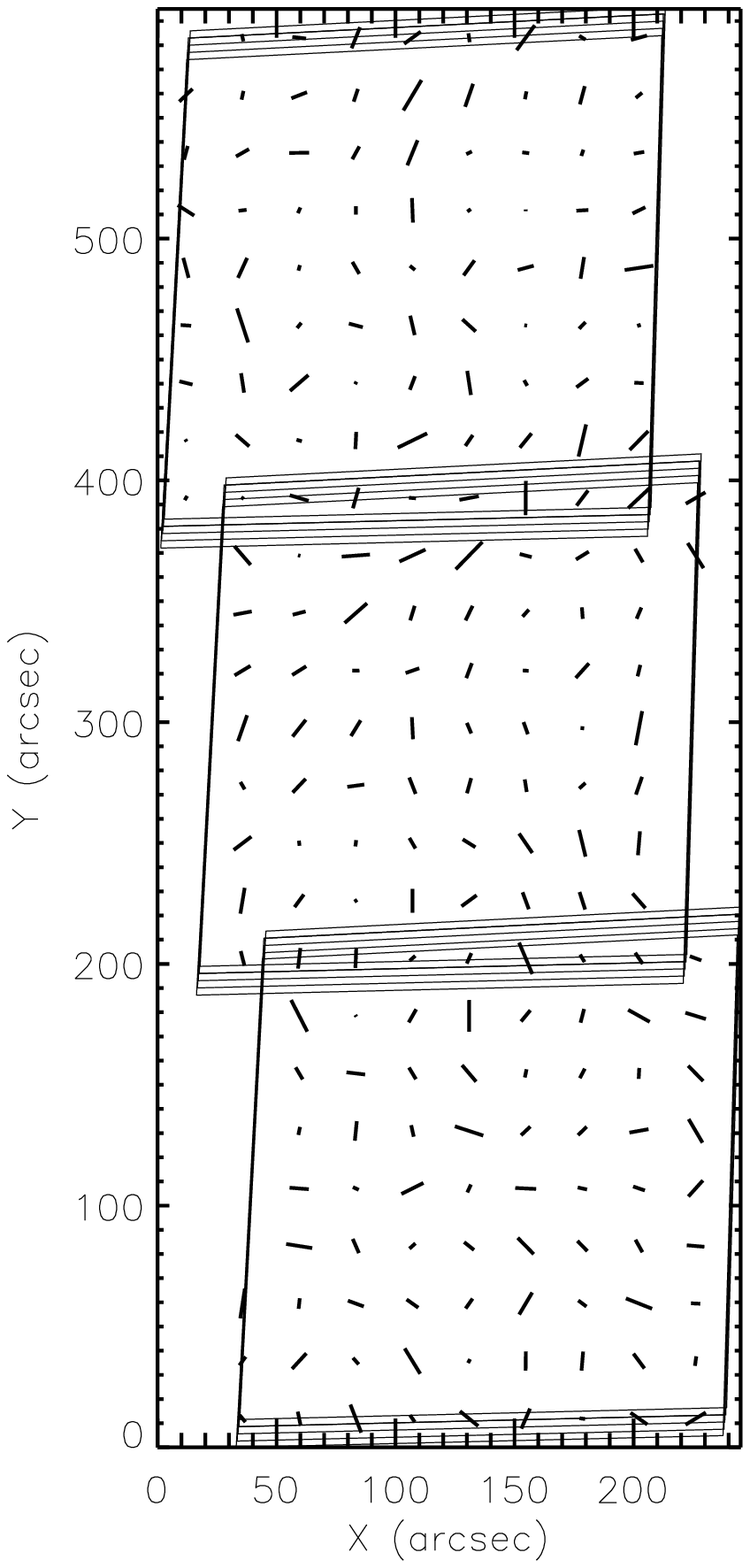}{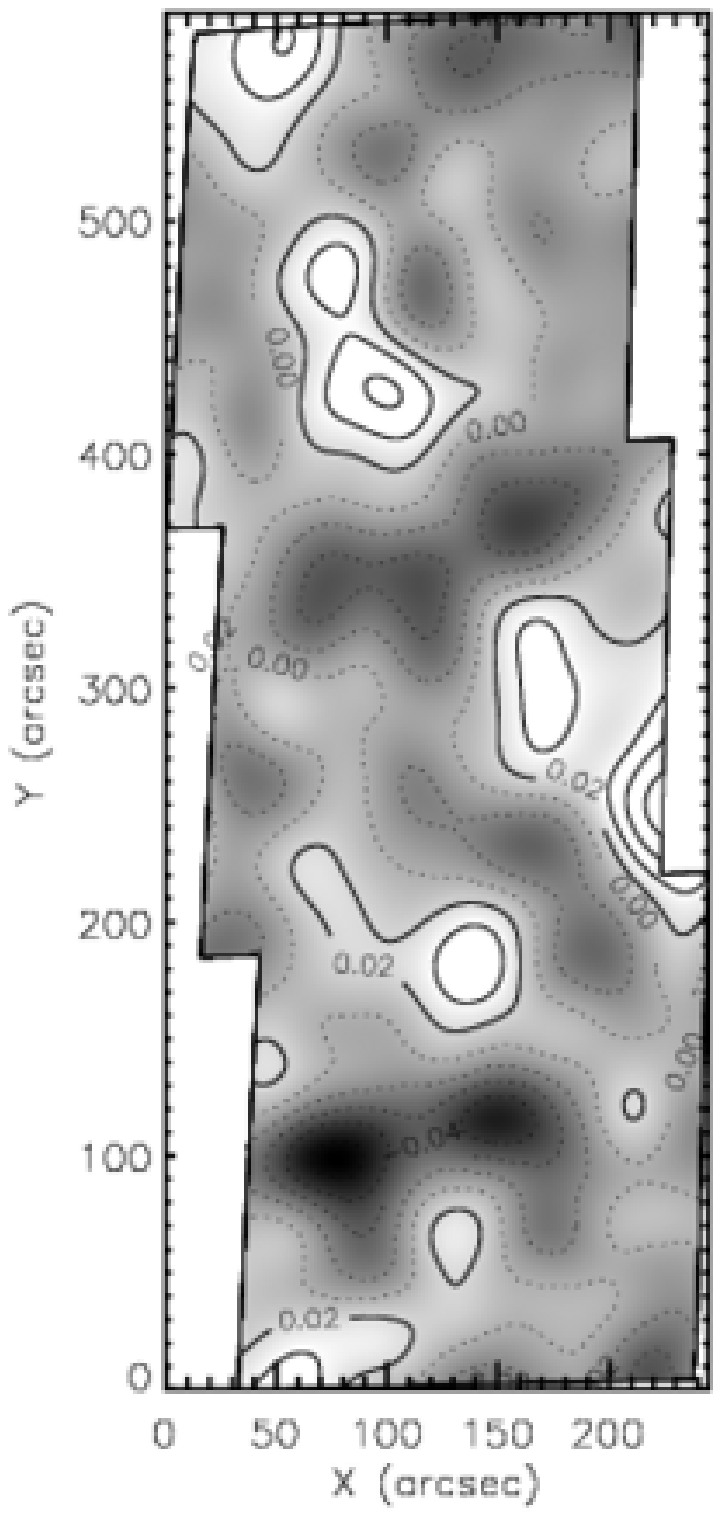}
\caption{Mass reconstruction in the Lynx field. The shear field (left panel)
is inverted to reconstruct the mass distribution (right panel).
This mass map has been smoothed with a FWHM$\sim40\arcsec$ Gaussian kernel.
The two high-redshift
clusters at $\bar{z}=1.265$ are clearly detected in good spatial
agreement with the cluster galaxies. In addition, the southern edge of the foreground cluster at $z=0.543$ is also visible
in this mass reconstruction though the cluster center is outside the current ACS pointings (see Figure~\ref{fig_xrayoverimage}).
The contours are spaced in $\delta\kappa=0.02$ interval. 
\label{fig_whisker}}
\end{figure}

We show the distortion and mass reconstruction of the Lynx field from this combined shape catalog in Figure~\ref{fig_whisker}.
Although the systematic alignments of source galaxies around the
cluster centers are subtle in the whisker plot (left panel), the resulting mass reconstruction (right panel)
clearly shows the dark matter concentration associated with the cluster galaxies. 
The mass map is generated using the maximum likelihood algorithm and is 
smoothed with a FWHM$\sim40\arcsec$ Gaussian kernel.
We verify that
other methods (e.g., Seitz \& Schneider 1995; Lombardi \& Bertin 1999) also produce
virtually identical results.

\begin{figure}
\plottwo{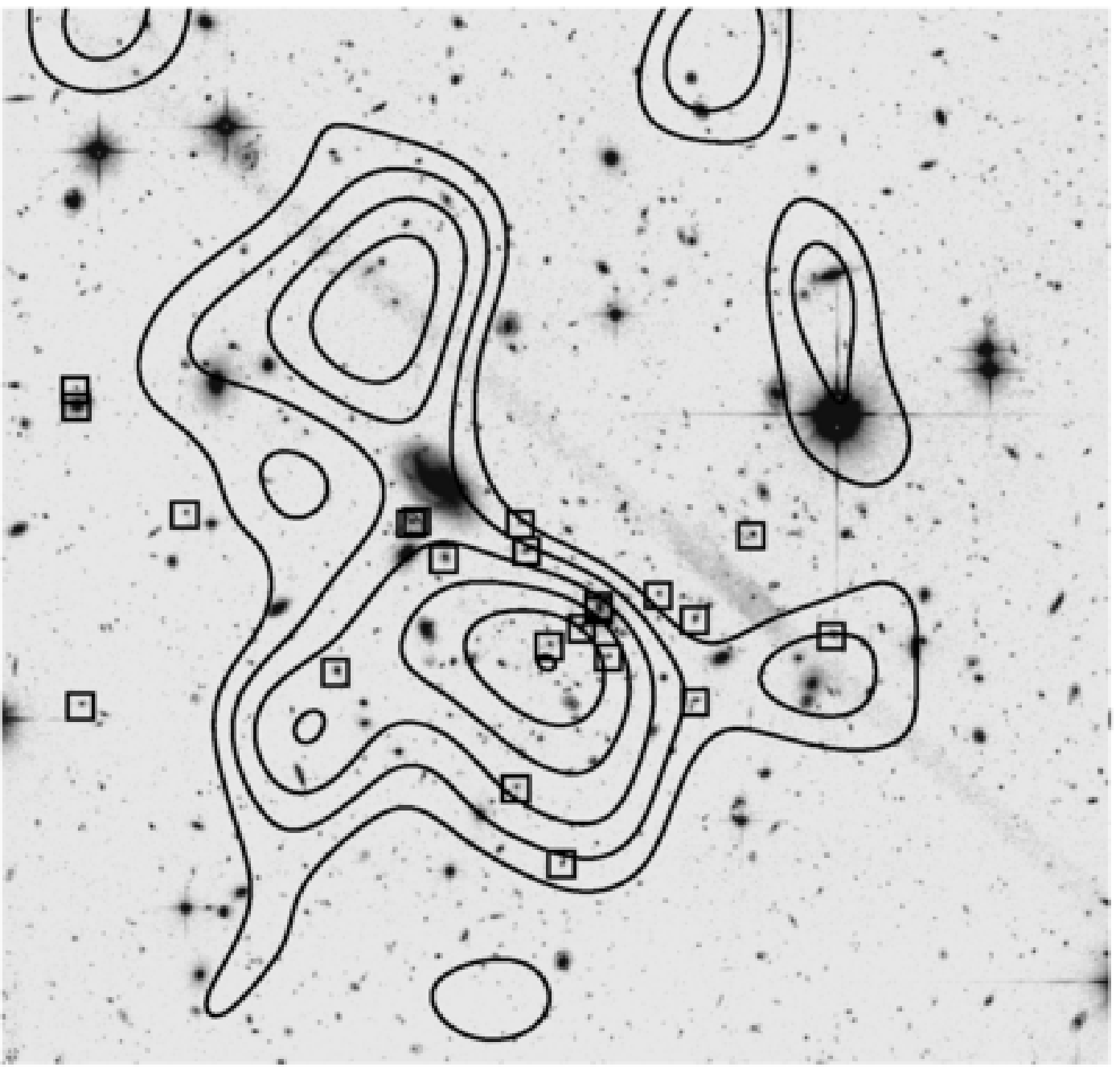}{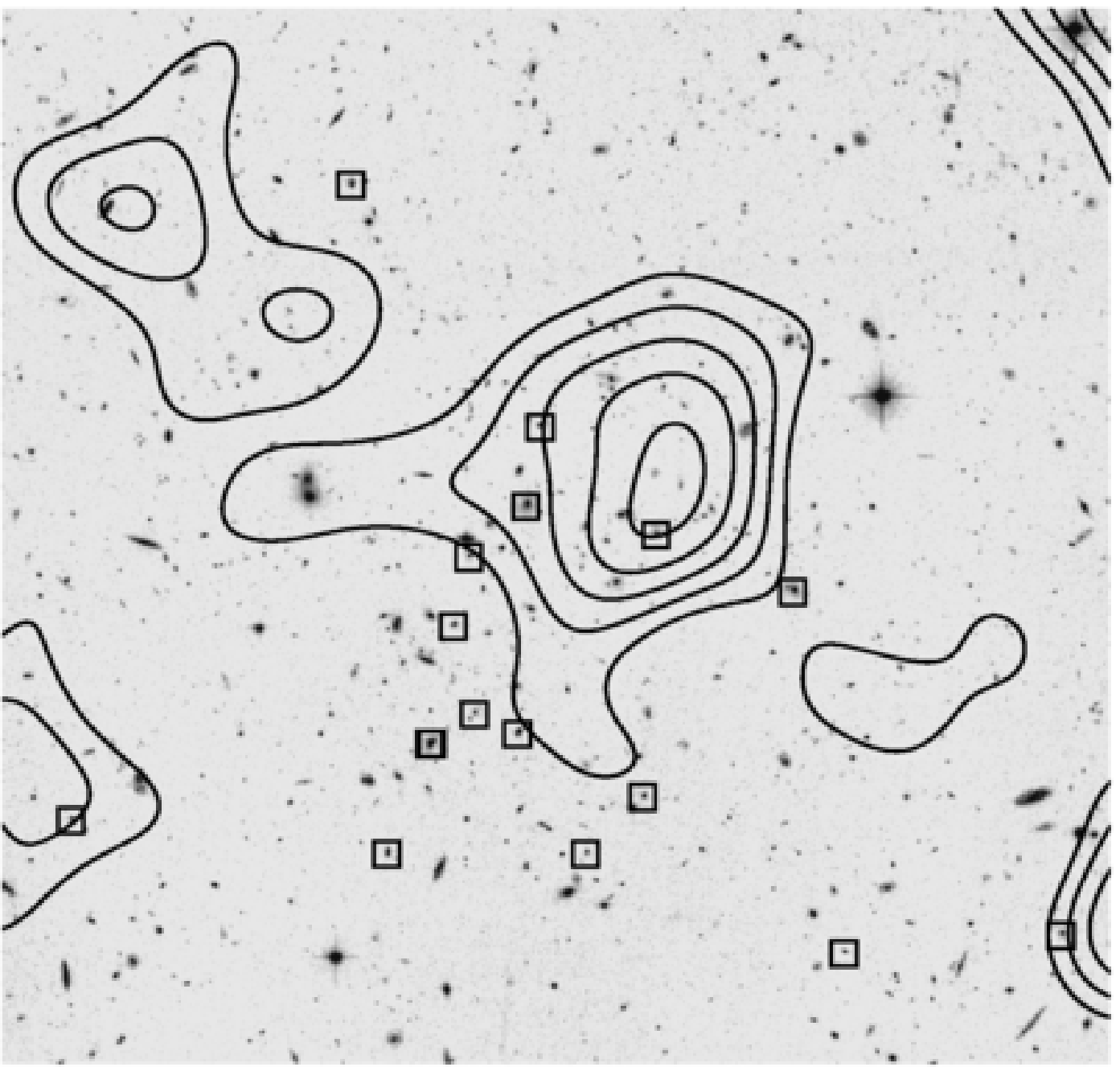}
\caption{High-resolution mass maps overlaid on top of the ACS image. We show the $\sim3\arcmin\times3\arcmin$
cuts of the Lynx-E and W regions on the left and right panels, respectively. The cluster red sequence candidates
(including the spectroscopic members)
are marked with square symbols.
The mass maps have been smoothed with
a FWHM$\sim20\arcsec$ Gaussian kernel. The clump associated with Lynx-E is $\sim10\arcsec$ offset from the 
BCGs and the Lynx-W clump seems to be centered on the western edge of the cluster galaxy distribution.
}
\label{fig_high_resolution}
\end{figure}

The two mass clumps are in good spatial agreement with both the cluster light and X-ray emission.
Within a radius of $1\arcmin$,
both clumps are found to be significant, above the $4 \sigma$ level (determined from bootstrap resampling).
Figure~\ref{fig_high_resolution} shows the high-resolution (smoothed with a FWHM$\sim20\arcsec$ kernel)
version of the mass maps overlaid on the ACS images. The clump associated with Lynx-E is offset $\sim10\arcsec$ from the 
BCGs and the Lynx-W clump seems to lie on the western edge of the cluster galaxy distribution.
In Paper I and II, we have reported significant mass-galaxy offsets for two clusters at $z\sim0.83$ 
and discussed the possibility that those offsets may signal the merging substructures.
Although it is tempting to interpret the mass-galaxy offsets in the current study as also
implying the similar merging of the two Lynx clusters, our investigation of 
the mass centroid distribution using the bootstrap resampling shows that the significance is only marginal
(i.e., the $r\sim10\arcsec$ circle roughly encloses $\sim70$\% of the centroid distribution).

It is encouraging to observe that the foreground cluster at $z\sim0.54$ affects 
the distortion of source galaxies and reveals itself 
in the weak lensing mass reconstruction (Figure~\ref{fig_whisker}) though
most of its galaxies are outside our ACS field (see Figure~\ref{fig_xrayoverimage}
for the location of the X-ray emission from the foreground cluster).
As shown by this foreground cluster and its manifestation in the mass map, light coming from
background galaxies is perturbed by all the objects lying in their paths to the observer.
Considering the high-redshifts ($\bar{z}=1.265$) of the Lynx clusters, the
likelihood of such interlopers is high. In addition, if the masses
of the two high-redshift clusters are not very large, even a moderately massive
foreground object can generate a similar lensing signal
because it has higher lensing efficiency for a fixed source plane (unless
the source plane is located at substantially higher than $z\sim1.3$).

\begin{figure}
\plotone{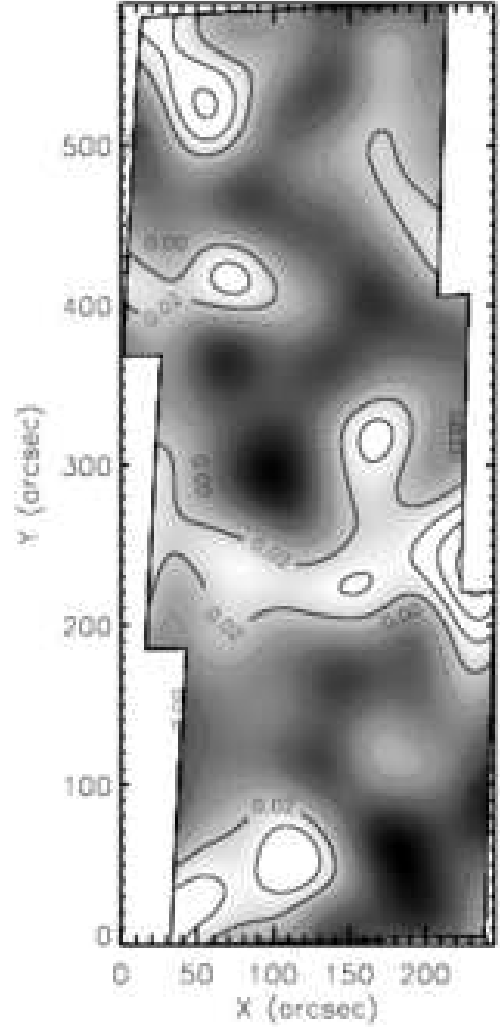}
\caption{Mass reconstruction using the bright ($22<z_{850}<25$) galaxies. We also
included the assumed cluster redsequence of the $\bar{z}=1.265$ clusters in this source sample.
It is remarkable to observe that the two high-redshift clusters disappear whereas
many of the other previous features in Figure~\ref{fig_whisker} still remain in this version.
}
\label{fig_mass_fore}
\end{figure}

In an attempt to separate this lower-redshift contribution from our weak lensing mass map
presented in Figure~\ref{fig_whisker}b, we created an alternate source sample
by selecting the brighter ($22<z_{850}<25$) galaxies. This time we did not exclude
the galaxies whose $i_{775}-z_{850}$ colors correspond to that of the cluster red-sequence because
they also serve as well-defined source plane at $z\sim1.3$ and their shapes should be 
perturbed by any lower-redshift mass clumps. We present this second
version of the mass reconstruction in Figure~\ref{fig_mass_fore}.
It is remarkable to observe that in this version
the two high-redshift clusters disappear whereas many of the assumed foreground
features (including the cluster at $z=0.54$) still remain.

The comparison of this second mass reconstruction with the previous result
also indicates that some of the foreground mass clumps might affect the
shape of the contours of the high-redshift clusters at large radii;
the mass clump of Lynx-E seems to have a neighboring foreground clump
at its southwestern edge, and the southern edge of the Lynx-W clump
also slightly touches the foreground structure (Figure~\ref{fig_high_resolution})
(However, far fewer galaxies were used for this second version of mass
reconstruction and thus the position of these structures have much less
significance). This apparent substructure in projection may bias our measurements of 
the total mass. We discuss this issue in \textsection\ref{section_mass_estimate}.
 
\subsection{Redshift Distribution of Source Galaxies}
As detailed in Paper I, the redshift distribution of the source galaxies of the Lynx field was inferred
from the photometric redshift catalog of the UDF. 
We also used the two photometric catalogs created from the Great Observatories Origins Deep
Survey (GOODS; Giavalisco et al. 2004) and the degraded UDF in order to estimate
the contamination of the cluster members in the source sample for $z_{850}<26$ and $z_{850}>26$, respectively.

\begin{figure}
\plotone{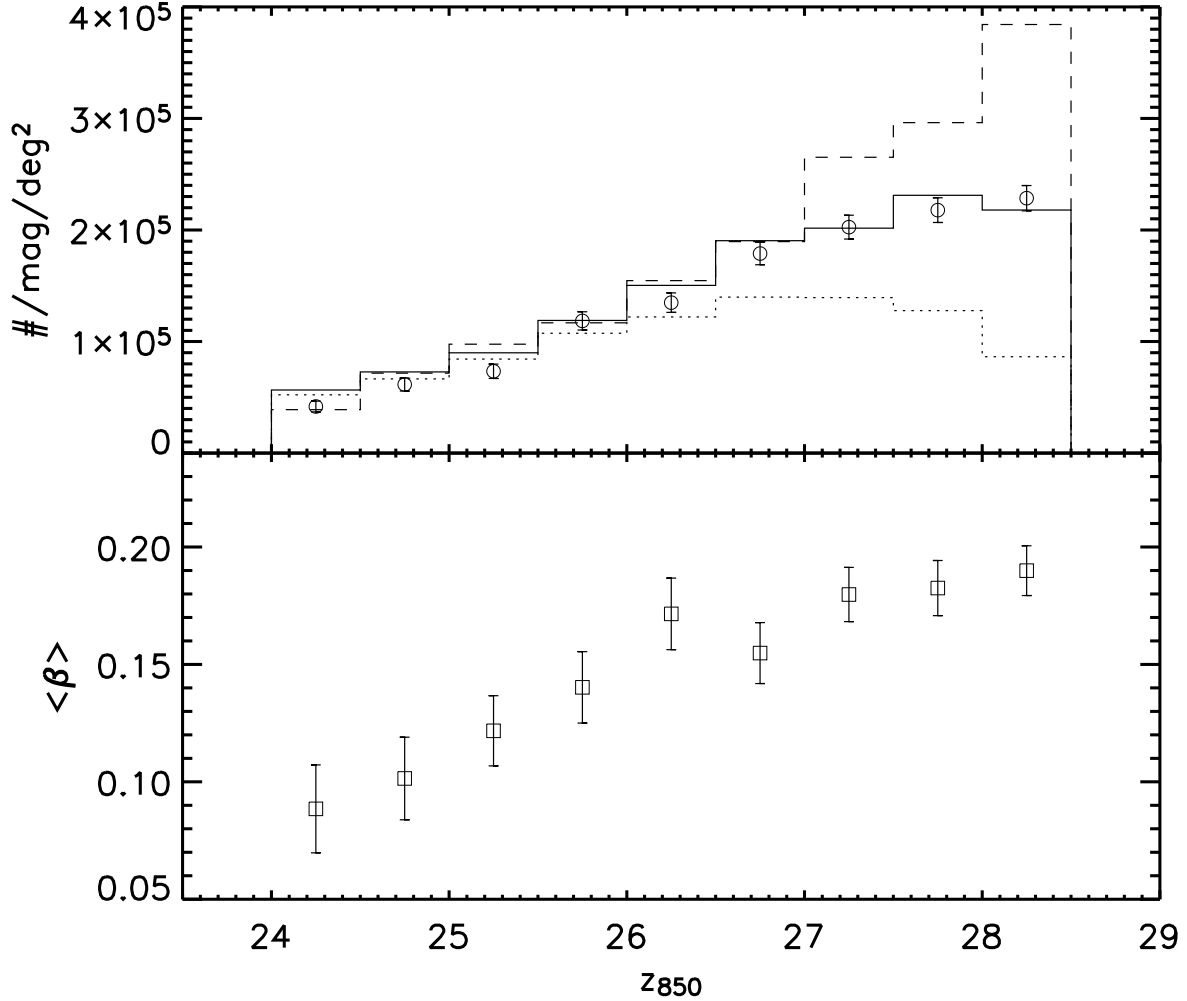}
\caption{Redshift distribution of source galaxies. The numbers
of galaxies per magnitude bin (normalized so that the number yields
the total number of galaxies per magnitude per square degrees)
in the Lynx, GOODS, and UDF are represented by
the solid, dotted and dashed lines (top panel). We degraded the UDF
images to match the S/N of the ACS Lynx data and thus
to estimate the contamination at $z_{850}>26$.
The redshift for each magnitude is calculated in terms
of the $\beta_{l}$ (see text for definition).
}
\label{fig_zdist}
\end{figure}

Figure~\ref{fig_zdist} shows the magnitude distribution of the source galaxies (top panel) with
the estimated mean redshift (bottom panel) for each magnitude bin. It appears that the number density excess
due to the cluster galaxy contamination is not significant throughout the entire magnitude range. However,
we must remember that the sample contains substantial contamination of foreground galaxies, which dilute the lensing signal. We
measure the mean redshift in terms of the following:

\begin{equation}
\beta_{l} = \left < \mbox{max} ( 0, \frac{D_{ls}} {D_s}) \right > \label{eqn_beta},
\end{equation}
\noindent
where $D_s$, $D_l$, and $D_{ls}$ are the angular diameter distance from the observer to the source,
from the observer to the lens and from the lens to the source, respectively.

We obtain $<\beta>=0.155$ for the entire source galaxies. The value corresponds to a single
source plane at $z_{eff}\simeq1.635$ and the critical surface mass density 
($\Sigma_c = c^2 (4 \pi G D_l \beta)^{-1}$) has the physical unit of $\sim6180 M_{\sun}/\mbox{pc}^2$
at the redshift of the lens $\bar{z}=1.265$. 

\subsection{Weak-lensing Mass Estimation \label{section_mass_estimate}}
\begin{figure}
\plottwo{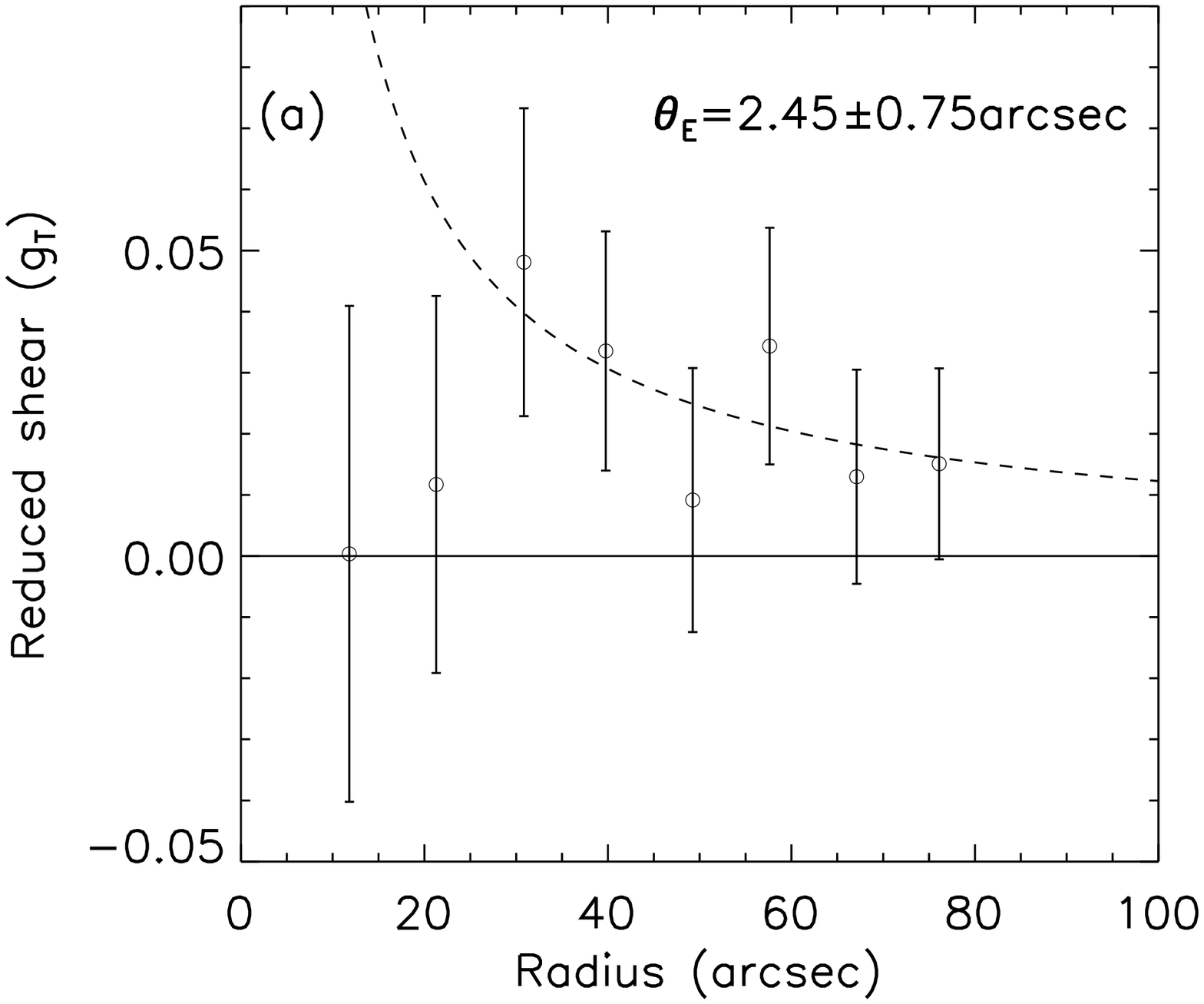}{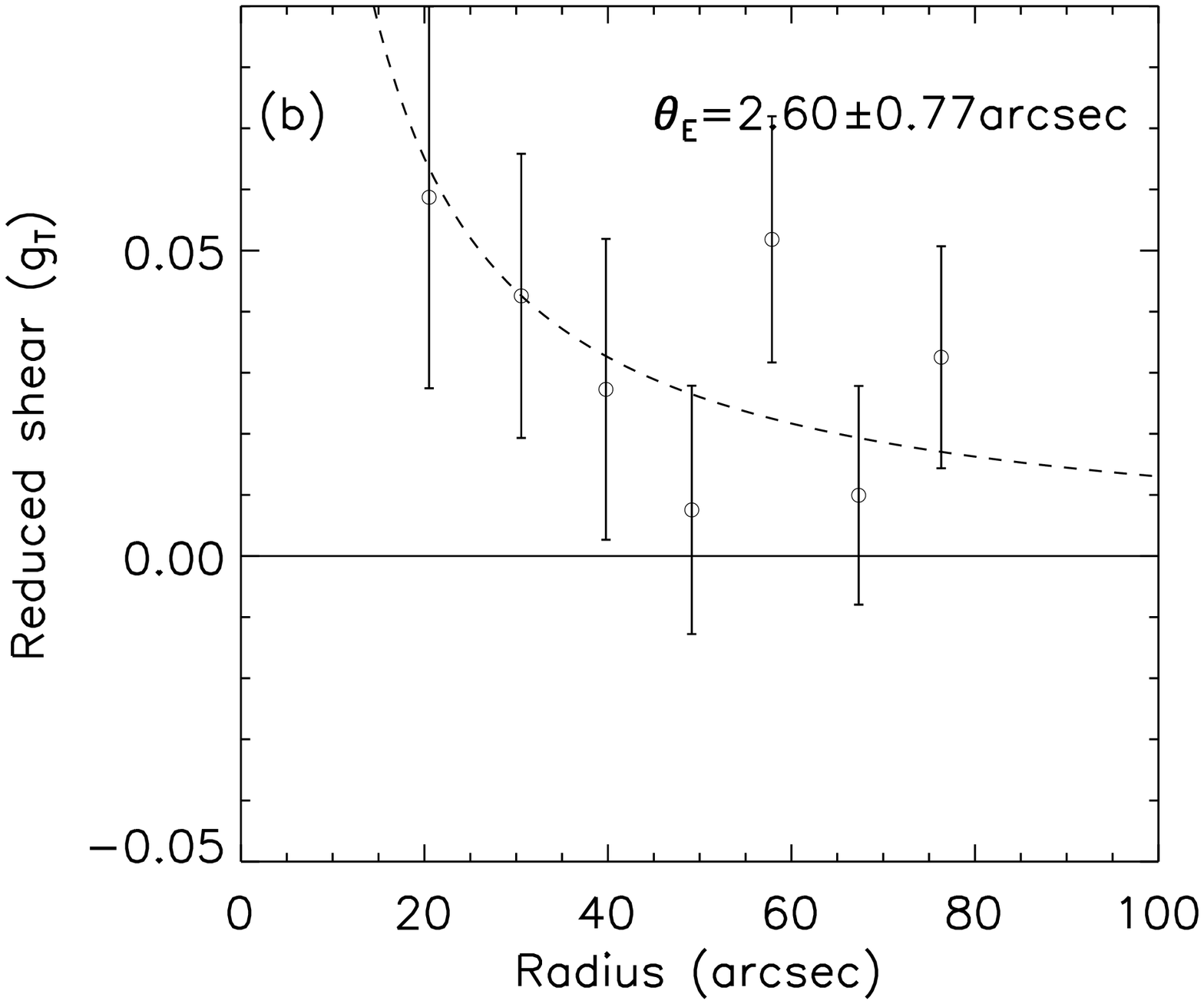}
\caption{Tangential (reduced) shear plots for the Lynx-E (a) and W (b) clusters.
Because of the proximity of the field boundary and the foreground
clusters, we limit the use of tangential shears to the values
taken at $r<85\arcsec$. We also excluded the $r<30\arcsec$ region
in order to avoid the low statistical significance and the possible substructure effects.
}
\label{fig_tan_shear}
\end{figure}

A first guess of the mass can be obtained by fitting the SIS model
to the observed tangential shears around the clusters. We chose
the origin of the tangential shears as the centroids
of the mass clumps in Figure~\ref{fig_whisker}. The neighboring
foreground structures at $z\simeq0.54$ as well as the
proximity of the field boundary restrict us to the use of
the tangential shears at radii no greater than $\sim80\arcsec$.
In addition, we discarded the measurements at $r<30\arcsec$
in order to minimize the possible substructure artifact and
the contamination of the lensing signal from the cluster members. 
Although this precaution leaves us with only a small fraction of
the total measurements, the lensing signal is clearly detected for both clusters
at the $\sim3\sigma$ level in the tangential shear plots (Figure~\ref{fig_tan_shear}).
It is plausible that the severly decreased shears at $r<30\arcsec$ for Lynx-E
might be in part caused by the aforementioned contamination from the cluster members.
We verified that the lensing signal disappeared when the background galaxies were
rotated by 45$\degr$ (null test).
Note that the uncertainties in Figure~\ref{fig_tan_shear} reflect only the statistical
errors set by the finite number of background galaxies. In Paper II, we demonstrated that the large scale structures
lying in front of and behind the high-redshift cluster MS 1054-0321 ($z\simeq0.83$) were dominant
source of errors in the mass determination, responsible
for approximately 15\% of the total cluster mass. This fractional uncertainty increases
substantially with cluster redshifts because the lensing by the foreground cosmic structures
become more efficient than the lensing by clusters whose redshifts approach those of source galaxies.
However, for the current clusters, we expect that the large statistical errors
still overwhelm the cosmic shear effects.
When we repeat the analysis of Paper II for the current clusters, we estimate that
the uncertainties of the Einstein Radius for the SIS fit marginally increases from
$\sigma_{er}=0\arcsec.75$ and $0\arcsec.77$ to
$\sigma_{er}=0\arcsec.81$ and $0\arcsec.83$ for Lynx-E and W, respectively.

The Einstein radius of $\theta_E=2\arcsec.45\pm0\arcsec.81$ (with respect to
the effective source plane at $z_{eff}\simeq1.635$)
for Lynx-E
corresponds to a mass of $M(r)=(4.0\pm1.3)\times10^{14} (r/\mbox{Mpc})~M_{\sun}$
and a velocity dispersion of $740_{-134}^{+113}\mbox{km}\mbox{s}^{-1}$.
Similar values of $M(r)=(4.2\pm1.4)\times10^{14} (r/\mbox{Mpc})~M_{\sun}$ and
$\sigma_{SIS}=762_{-133}^{+113}\mbox{km}\mbox{s}^{-1}$
are obtained for Lynx-W as implied by its comparable
Einstein Radius $\theta_E=2\arcsec.60\pm0\arcsec.83$.

As mentioned in \textsection\ref{subsection_acsobservation}, we note that there is
a strongly lensed arc candidate at $r\simeq4.5 \arcsec$ for Lynx-E, which
can provide a useful consistency check.
In general,
Einstein radii depend on source redshifts, and the relation steepens if a lens
is at a high redshift. If the Einstein radius of the arc is assumed to be $\theta_E=4.5\arcsec$,
this implies that the redshift of the object should lie at $1.8<z<3.2$ in our adopted cosmology
(the uncertainty reflects only the errors of the Einstein radius from the SIS fit result).
Because we have only $i_{775}$ and $z_{850}$ band images, the photometric redshift estimation
of this arc candidate is unstable. Nevertheless, if we use the HDFN prior and truncate
it below $z=1.2$, the color ($i_{775}-z_{850}=0.098$) of the object is consistent
with the SED of the starburst galaxy at $1.7<z<3.7$.

\begin{figure}
\plotone{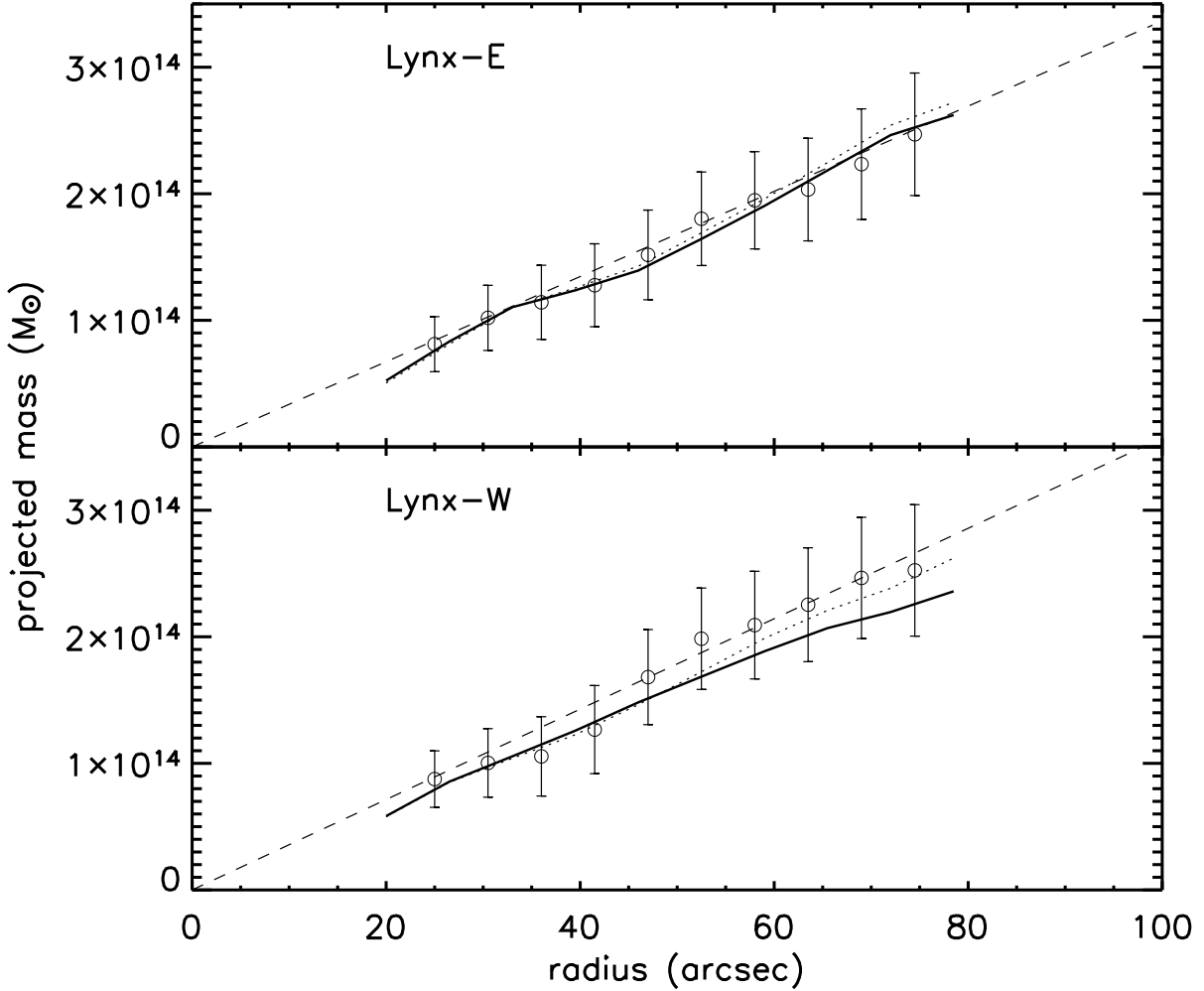}
\caption{Cumulative projected cluster mass versus radius for
Lynx-E (top) and W (bottom). The aperture densitometry (open circles)
yields very similar masses obtained 
from the rescaled mass map (dotted) and the
SIS fit (dashed). The solid lines represent the cluster masses
estimated when the foreground structure is absent. It seems that
both clusters are not significantly affected due to these
intervening mass clumps.
Within 0.5 Mpc ($\sim60\arcsec$) aperture radii, Lynx-E and W
enclose a similar mass of ($2.0\pm0.5) \times 10^{14} M_{\sun}$.
}
\label{fig_mass_summary}
\end{figure}

Alternatively, we can also estimate the cluster mass
based on the two parameter-free methods, namely the aperture mass densitometry and the rescaled mass
reconstruction. Although these two parameter-free approaches need some feedbacks from the above SIS fitting result 
to lift the mass-sheet degeneracy, in general they provide
more robust methodology. They are less affected by the cluster substructure or the deviation
from the assumed radial profile. However, one drawback of this approach is that
the measurement is more severely influenced by the cosmic shear effect than
in the case of the SIS fitting because the aperture mass densitometry uses less amount
of information (i.e., decreased tangential shears in outer range).
With the $r=80-90\arcsec$ region as a control annulus for both clusters,
we computed the cluster mass profiles from these two parameter-free methods (Figure~\ref{fig_mass_summary}); 
from the SIS fit results, we determine the mean mass density in the annulus
to be $\bar{\kappa}=0.014\pm0.004$ and $0.015\pm0.005$ for Lynx-E and W, respectively.
As observed in Paper I and II, the 
mass estimation obtained
from the rescaled mass reconstruction (dotted) is in good agreement with the aperture mass densitometry (open
circle). We also note that both methods gives masses consistent with the SIS fit results.

Because we used the SIS fit results above to lift the mass-sheet degeneracy, it is useful to
examine how the result change when an NFW profile is assumed, instead. Unfortunately, the low lensing signal
in the limited range does not allow us to constrain the two free parameters of the NFW profile simultaneously;
the two parameters trade off with each other without significantly altering the quality of the fit.
Freezing the concentration parameter to $c=4$, nevertheless, yields $r_s=180\pm37$ ($187\pm34$) kpc for
Lynx-E (W), predicting the mean mass density of $\bar{\kappa}=0.015\pm0.020$ ($0.016\pm0.021$) in the control
annulus. Different choices for the concentration parameter $c$ do not change these results substantially
(for instance, the choice of $c=6$ gives $\bar{\kappa}\simeq0.012$ for Lynx-E). 

In \textsection\ref{section_mass_reconstruction} we demonstrated that
both clusters might have neighboring foreground mass clumps in projection. Therefore, it is
worthwhile to assess how much these foreground structures affect our mass estimation.
Because the redshift information of the foreground masses are not available, we
cannot subtract their contribution directly from our mass map.
Instead, we attempted to minimize their effects by replacing the mass density
of the region that is occupied by the foreground mass clumps
with the azimuthal average from the rest. Of course, we do not expect that this scheme 
yields cluster masses that are completely free from foreground contamination, since
the azimuthal averages taken at other regions might be biased. 
However, this method is still an important test because
a significant difference in resulting mass estimation must be detected if the foreground
contamination is indeed severe.

The mass-sheet lifted mass map is convenient for this type of analysis. We replaced
the southwestern region ($\sim220\degr<\theta<\sim260\degr$; the angle is measured from
the north axis counterclockwise) of the Lynx-E clump and the southern
region ($\sim130\degr<\theta<\sim195\degr$) of the Lynx-W clump with
the azimuthal averages taken at different angles.
The solid lines in Figure~\ref{fig_mass_summary} represent the mass profiles
obtained from this measurement. For both clusters, this new measurements
give slightly lower values, but the change is only marginal.
We estimate that both Lynx-E and W have 
a similar mass of $(2.0\pm0.5) \times 10^{14} M_{\sun}$ within
0.5 Mpc ($\sim60\arcsec$) aperture radius from this approach.
The uncertainties here are estimated from 5000 bootstrap
resampling of the source galaxies and we do not include the cosmic
shear effects because it is non-trivial to estimate the effect
for this rescaled mass map approach. 

We adopt the conventional definition of the virial radius, where
the enclosed mean density within the sphere becomes 
200 times the critical density $\rho_c(z)=3H(z)^2/8\pi G$ at the redshift of the cluster.
Although the factor 200 above is most meaningful in the mass-dominated flat universe,
we retain this definition so as to enable 
a consistent comparison with the values of other clusters found in the literature.
The assumption of the spherical symmetry (SIS)
allows us to estimate $r_{200}\simeq0.75$~Mpc 
and $M_{200}\simeq2.0\times 10^{14} M_{\sun}$ for both Lynx clusters.
These virial properties are much smaller than the clusters at $z\sim0.83$
studied in Paper I and II. We reported that CL 0152-1357 has a virial radius of $r_{200}\sim1.1$~Mpc
and a virial mass of $M_{200}\sim4.5\times10^{14}~M_{\sun}$ in Paper I. For
MS 1054-0321, Paper II quoted $r_{200}\simeq1.5$~Mpc and $M_{200}\simeq1.1\times10^{15}~M_{\sun}$.
If we assume that the two Lynx clusters are approaching each other perpendicular to the line of sight at a free-fall speed,
our order-of-magnitude estimation predicts that 
the two Lynx cluster will merge into a single cluster whose virial
mass exceeds
$\sim 4.0\times 10^{14} M_{\sun}$ in a time scale of $t\sim2$~Gyrs (or at $z\sim0.8$). 

\section{$CHANDRA$ X-RAY ANALYSIS}
\subsection{Cluster Temperature and Luminosity \label{section_temperature}}
The X-ray spectra of Lynx-E and Lynx-W were extracted from the circular regions ($\bar{r}\sim36\arcsec$) positioned
at their approximate X-ray centroids after the point sources (Stern et al. 2002) are
removed. The redistribution matrix file (RMF) and the area response file (ARF) were created using the CIAO tool version 3.2 
with the calibration database (CALDB) version 3.1, which
properly accounts for the time-dependent low-energy QE degradation, as well as charge transfer inefficiency (CTI).
The photon statistics is somewhat poor mainly because the clusters are at a high-redshift ($\bar{z}=1.265$) and 
thus the differential surface brightness dimming is severe $\sim (1+z)^4$. Especially, as implied by its low
temperature ($T<2$ keV), the Poissonian scatter of the Lynx-W is worse. Therefore, we constructed the spectra
for both clusters with a minimum count of 40 per spectral bin. We think that this choice makes the
spectral fitting stable without diluting the overall shape of photon distribution too much. Because it
is impossible to constrain the iron abundance given the statistics, we fixed the metallicity at 0.36 $Z_{\sun}$.
This assumes that both Lynx clusters possess similar metallicity to RDCS 1252.9-2927 at $z=1.24$ (Rosati et al. 2004).
However, as noted by Stanford et al. (2001), we observed only minor changes even when different values were tried.
The Galactic hydrogen column density was also fixed at $\mbox{n}_H=2.0\times10^{20}\mbox{cm}^{-2}$ (Dickey \& Lockman 1990).
We used the $\chi^2$ minimization modified by Gehrels (1986;CHI-GEHRELS), who extended the conventional
$\chi^2$ statistics so that it can handle the deviation of the Poissonian from the Gaussian at
the low-count limit. 

\begin{figure}
\plottwo{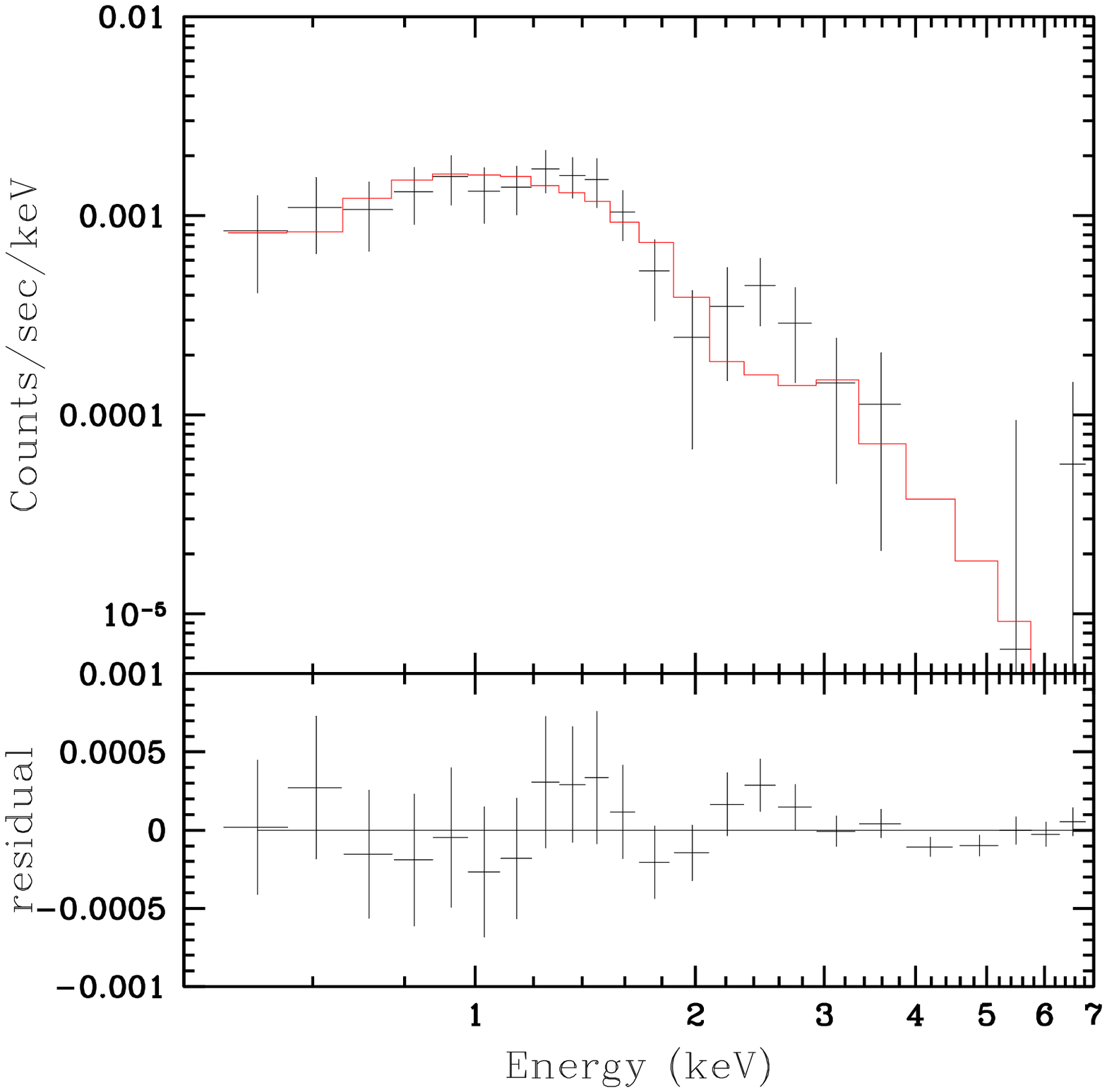}{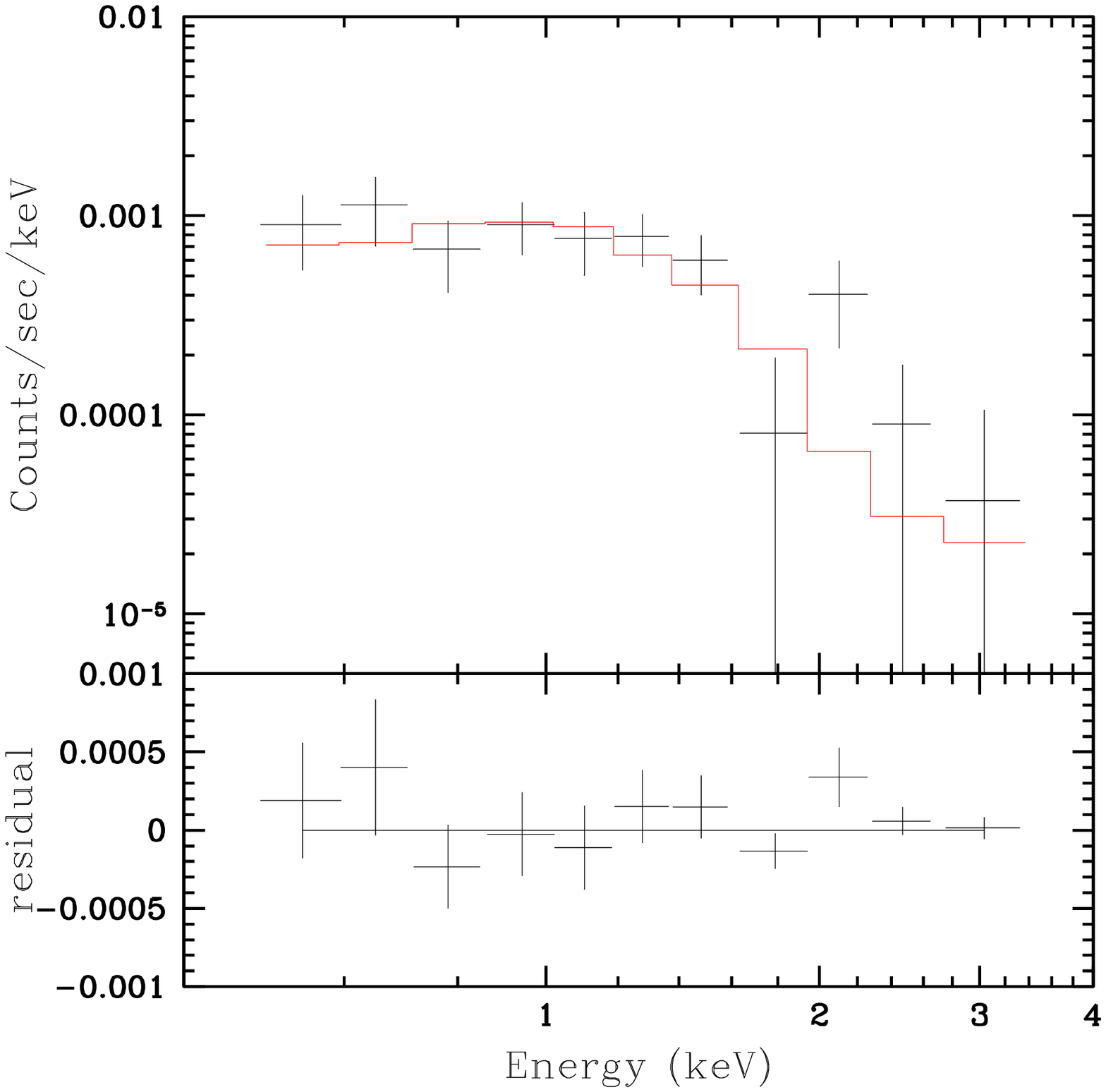}
\caption{X-ray spectra with best-fit MEKAL model. We estimate
that a temperature of $T=3.8_{-0.7}^{+1.3}$~keV for Lynx-E (left panel) 
and $T=1.7_{-0.4}^{+0.7}$~keV for Lynx-W (right panel). We froze
the abundance, redshift and column density at $Z=0.36Z_{\sun}$,
$z=1.26$ (1.27 for Lynx-W), and $\mbox{n}_H=2.0\times10^{20}\mbox{cm}^{-2}$ (Dickey \& Lockman 1990),
respectively.
}
\label{fig_spec}
\end{figure}

Figure~\ref{fig_spec} shows the best-fit MEKAL plasma spectra (Kaastra \& Mewe 1993; Liedahl, Osterheld, \& 
Goldstein 1995) for both clusters. We obtain $T=3.8_{-0.7}^{+1.3}$~keV for Lynx-E with a
reduced $\chi^2$ of 0.79 (19 degrees of freedom). Lynx-W is determined to have $T=1.7_{-0.4}^{+0.7}$~keV
with a reduced $\chi^2$ of 1.16 (7 degrees of freedom). 

The observed fluxes are estimated to be $F (0.4-7\mbox{keV})=1.5_{-0.2}^{+0.3}\times10^{-14}\mbox{ergs~cm}^{-2}~\mbox{s}^{-1}$ and
$F (0.4-4\mbox{keV})=7.2_{-0.5}^{+1.4}\times10^{-15}\mbox{ergs}~\mbox{cm}^{-2}~\mbox{s}^{-1}$, which
can be transformed into the rest-frame (also, apertured-corrected to $\sim0.5$~Mpc) bolometric (0.01 - 40 keV) 
luminosity of $L_X=(2.1\pm0.5)\times10^{44}$ 
and $(1.5\pm0.8)\times10^{44}~\mbox{ergs}~\mbox{s}^{-1}$ for Lynx-E and Lynx-W, respectively
(note that the shallow surface brightness profile of Lynx-W requires a rather large aperture
correction factor).

\subsection{X-ray Surface Brightness Profile and Mass Determination \label{section_sb}}
\begin{figure}
\plottwo{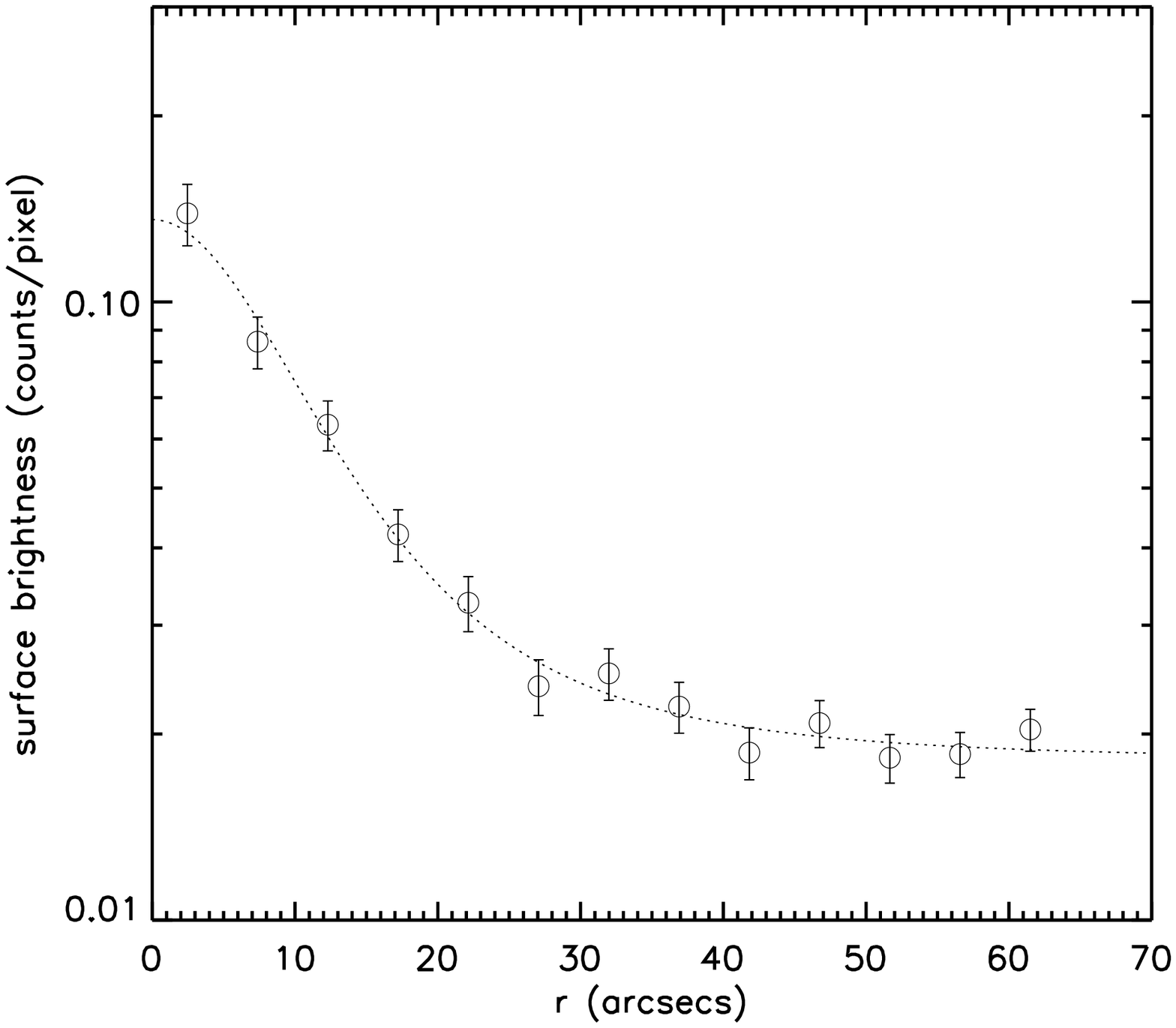}{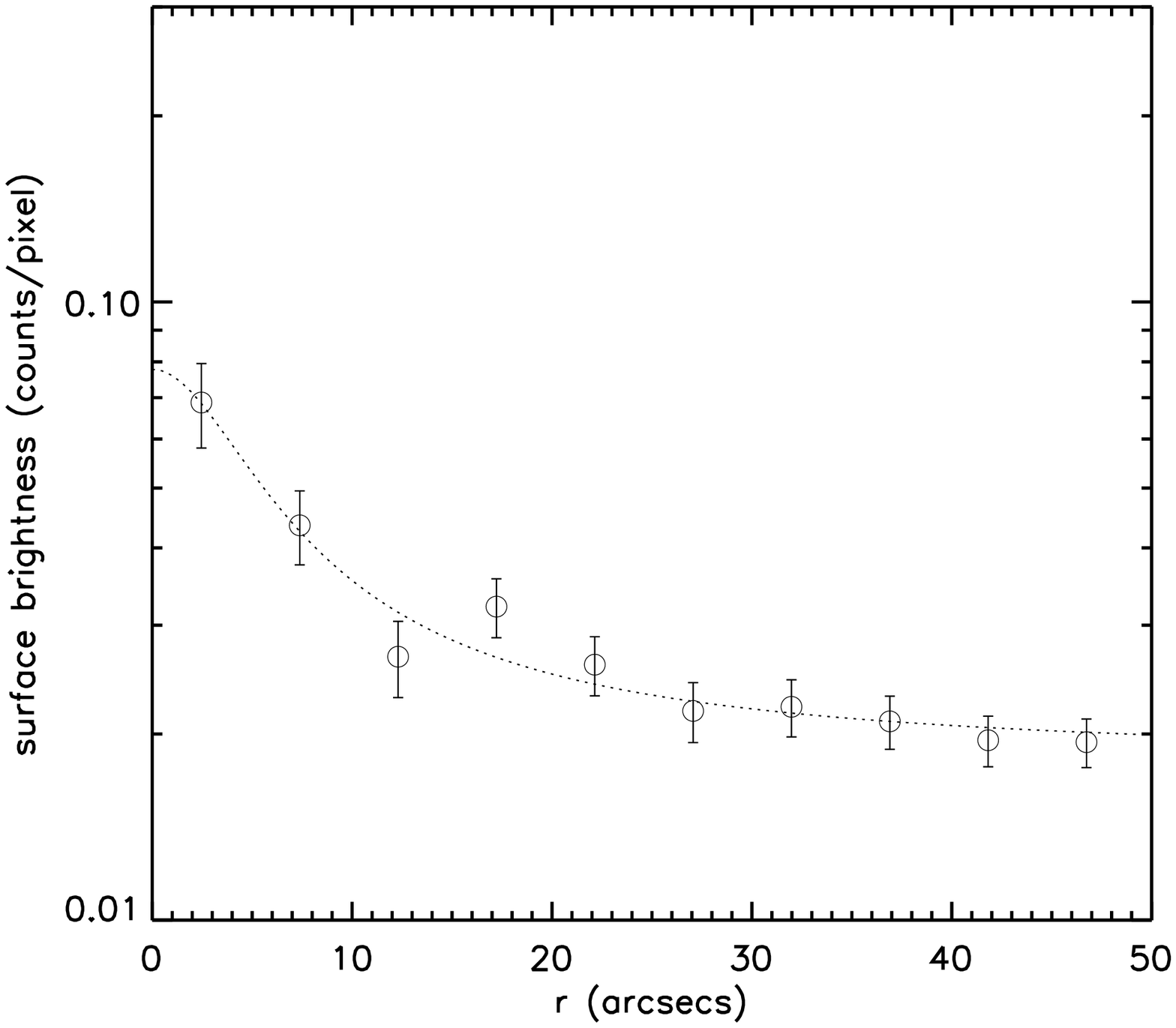}
\caption{Surface brightness fit. We obtain $\beta=0.71\pm0.12$ and $r_c=13.2\pm3.2$
for Lynx-E (left panel), and $\beta=0.42\pm0.07$ and $r_c=4.9\arcsec\pm2.8\arcsec$
for Lyny-W (right panel).
}
\label{fig_betafit}
\end{figure}

The azimuthally averaged radial profiles were created from the exposure-corrected $Chandra$ image.
In Figure~\ref{fig_betafit} we display these radial profiles with the best-fit isothermal beta models
for both clusters. As is indicated by their X-ray image and cluster galaxy distribution, 
Lynx-E has a higher concentration ($\beta=0.71\pm0.12$ and $r_c=13.2\pm3.2$) of the ICM
than Lynx-W ($\beta=0.42\pm0.07$ and $r_c=4.9\arcsec\pm2.8\arcsec$). 

Together with the cluster temperatures determined in \textsection\ref{section_temperature}, these
structural parameters can be converted to the cluster mass under the assumption of hydrostatic
equilibrium. In general, many authors report cluster masses within a spherical volume rather
than a cylindrical volume spanning from the observer to the source plane, which however
is the preferred and natural choice in weak-lensing measurements. This different
geometry is often a source of confusion and subtlety in mass comparison between
both approaches. Therefore, in this paper we present our X-ray mass estimates in a cylindrical volume in order
to ensure more straightforward comparison with the weak-lensing result using the following equation (Paper II):

\begin{equation}
M_{ap}(r)= 1.78  \times 10^{14} \beta \left ( \frac{T}{\mbox{keV}} \right ) 
\left (  \frac{r}{\mbox{Mpc}}  \right )  \frac{r/r_c}{\sqrt{1+(r/r_c)^2}} M_{\sun} \label{eqn_xray_mass_2d}
\end{equation}
\noindent

For Lynx-E we obtain $M(r\leq0.5~\mbox{Mpc})=2.3_{-0.4}^{+0.8}\times10^{14} M_{\sun}$, which is in
good agreement with our weak-lensing measurement. 
On the other hand, the X-ray mass of Lynx-W ($M(r\leq0.5~\mbox{Mpc})=6.3_{-1.5}^{+2.6}\times10^{13} M_{\sun}$)
is much lower than the weak lensing estimation. We will discuss a few possible scenarios for this discrepancy
in \textsection\ref{summary}.

\section{COMPARISON WITH OTHER STUDIES}
The first attempt to estimate the mass of Lynx-W was made by Stanford et al. (1997)
using the X-ray luminosity from the ROSAT-PSPC observation and
the velocity dispersion obtained from the Keck spectroscopy of 8 galaxies.
They converted the luminosity $L_X\sim1.5\times10^{44} \mbox{ergs}~\mbox{s}^{-1}$ to
$M(r<2.3~ \mbox{Mpc})\sim7.8\times10^{14} M_{\sun}$ assuming $\beta=0.8$. A similar
value of $M(r<2.3 \mbox{Mpc})=5.4_{-2.3}^{+3.1}\times10^{14} M_{\sun}$ was estimated
from the velocity dispersion of $\sigma=700\pm180 \mbox{km}~\mbox{s}^{-1}$ (note
that they adopted $h_{100}=0.65$ and $q_0=0.1$).
Although both masses are consistent with each other, 
their X-ray luminosity measurement seems to have suffered a severe contamination
from the neighboring point sources, which are now identified in the $Chandra$ observation.

In their presentation of the $Chandra$ analysis,
Stanford et al. (2001) did not attempt to estimate the mass of Lynx-W
because of the large uncertainty of the temperature measurement, as well as the apparent
asymmetry of the X-ray emission. 
Our predicted velocity dispersion of $\sigma_{SIS}=762_{-133}^{+113}\mbox{km}\mbox{s}^{-1}$
from the SIS fit result is consistent with their most recent determination 
of the velocity dispersion $\sigma=650\pm170~\mbox{km}~\mbox{s}^{-1}$ from the spectroscopic
redshifts of the 9 member galaxies. Lynx-W was also selected as one of the 28 X-ray clusters for
the study of the X-ray scaling relation at high redshifts by Ettori et al. (2004). From the
re-analysis of the $Chandra$ data, they obtained $\beta=0.97\pm0.43$, $r_c=163\pm70$~kpc, and $T_X=2.9\pm0.8$~keV, which
predicts a projected mass of $M(r\leq 0.5~\mbox{Mpc})= 3.0\pm1.5\times10^{14} M_{\sun}$ (eqn.~\ref{eqn_xray_mass_2d}).
This mass is consistent with our weak-lensing estimation ($2.0\pm0.5) \times 10^{14} M_{\sun}$, but
much higher than the value from our re-analysis of the same $Chandra$ data ($6.3_{-1.5}^{+2.6}\times10^{13} M_{\sun}$).
In general, many detailed steps in the $Chandra$ X-ray analysis such as the QE correction, background
modeling, flare removal, spectral aperture, etc. affect the final result, and much more
if the source is faint. Therefore, it is difficult, if not impossible, to trace the exact causes
of the differences. Nevertheless, we note that there is an important difference in the calibration
of the low-energy quantum efficiency correction between the results. Ettori et al. (2004) used
the ACISABS correction method (Chartas and Getman 2002) to account for the low-energy QE degradation, which is however 
now officially disapproved by the $Chandra$ $Data$ $Center$. We also demonstrate in Paper II that
the use of this ACISABS model causes a difference of $\sim1$~keV in the temperature determination of MS1054-0321.
We suspect that the effect should be more important in Lynx-W because of
its low temperature and luminosity.

Stanford et al. (2001) obtained an X-ray temperature of $5.8_{-1.7}^{+2.8}$ keV
for Lynx-E. Combined with their determination of $\beta=0.61\pm0.12$ and
$r_c=11\arcsec.14\pm3\arcsec.41$, this gives
a projected mass of $M(r\leq0.5\mbox{Mpc})=3.1_{-0.9}^{+2.4} \times 10^{14} M_{\sun}$ 
(eqn.~\ref{eqn_xray_mass_2d}), which is slightly higher
than our X-ray re-analysis of the same $Chandra$ data by $\sim35$\% though the error bars from
both results marginally overlap. Vihklinin et al. (2002) included Lynx-E in their
sample of the 22 distant clusters to study the evolution the X-ray scaling relation.
With the early understanding of the low-energy QE problem of the $Chandra$, they obtained
$T_X=4.7\pm1.0$~keV, $r_c=167$~kpc, and $\beta=0.85\pm0.33$. Using the ACISABS correction. Ettori et al. (2004)
reported $T_X=5.2_{-1.1}^{+1.6}$~keV, $r_c=128\pm40$~kpc, and $\beta=0.77\pm0.19$.
The results from these two papers are statistically consistent with, but 
slightly higher than our values ($T=3.8_{-0.7}^{+1.3}$keV, $r_c=111\pm27$, and $\beta=0.71\pm0.12$),
which predicts the lowest projected mass of $M(r\leq0.5\mbox{Mpc})=2.3_{-0.4}^{+0.8}\times10^{14} M_{\sun}$.
As already mentioned in the discussion of the Lynx-W temperature above, we suspect that
the difference in temperatures mainly stems from the different correction methods of the low-energy
QE degradation. 
Although our understanding of the $Chandra$ instrument still evolves and this may
neccesitate some updates to our results,
it is encouraging to
note that this X-ray mass is closest to
our independent lensing determination of the cluster mass of
$M(r\leq0.5\mbox{Mpc})=(2.0\pm0.6)\times10^{14} M_{\sun}$ from the SIS fit result.
Our spectroscopic catalog currently provides the redshifts of 11 member galaxies
within a $r=80\arcsec$ radius (B. Holden in prep). Based on Tukey's biweight estimator,
we obtain a velocity dispersion of $720\pm140 \mbox{km}~\mbox{s}^{-1}$ (without
assuming a Gaussian distribution). This direct measurement agrees with the predicted
velocity dispersion of $740_{-134}^{+113}\mbox{km}\mbox{s}^{-1}$ from the lensing
analysis (\textsection\ref{section_mass_estimate}). In addition,
the cluster temperature $T_X=3.8_{-0.7}^{+1.3}$keV with $\beta=0.71$ is translated into
$\sigma_v=662_{-64}^{+106} \mbox{km}\mbox{s}^{-1}$ (from $\beta=\mu m_p \sigma_v^2/kT_X$),
in good agreement with both results.

\section{DISCUSSION AND CONCLUSIONS \label{summary}}
We have presented a weak-lensing analysis of the two Lynx clusters at $\bar{z}=1.265$
using the deep ACS $i_{775}$ and $z_{850}$ images. Our mass reconstruction
clearly detects the dark matter clumps associated with the two high-redshift
clusters and other intervening objects within the ACS field, including
the known foreground cluster at $z=0.57$.
In order to verify the significance of the cluster detection and
to separate the high-redshift signal from the low-redshift contributions,
we performed a weak-lensing tomography by selecting an alternate
lower-redshift source plane. This second mass reconstruction does not
show the mass clumps around the high-redshift clusters, while maintaining
most of the other structures seen in the first mass map. This experiment
strongly confirms that the weak-lensing signals observed in the
first mass reconstruction are real and come from the high-redshift Lynx clusters.

Interestingly, both clusters are found to have similar weak-lensing masses
of $\sim 2.0\times 10^{14} M_{\sun}$ within 0.5 Mpc ($\sim60\arcsec$) aperture radius
despite their discrepant X-ray properties. Our re-analysis of the Chandra archival data
with the use of the latest calibration of the low-energy QE degradation
shows that Lynx-E and W have temperatures of $T=3.8_{-0.7}^{+1.3}$ and
$1.7_{-0.4}^{+0.7}$~keV, respectively.
Combined with the X-ray surface brightness profile measurements, the X-ray temperature of Lynx-E
gives a mass estimate in good agreement with the weak-lensing result. On the other hand,
the X-ray mass of Lynx-W is much smaller than the weak-lensing estimation nearly by a factor of three.
According to our experiment in \textsection\ref{section_mass_estimate},
it is unlikely that any foreground contamination or cosmic shear effect in
weak-lensing measurement causes this large discrepancy. 

Apart from a simplistic, but valid possibility that Lynx-W might have a filamentary structure extended along the line of sight,
yielding a substantial, projected mass but with yet only low-temperature thermal emission, we can also
consider the self-similarity breaking (e.g., Ponman et al. 1999; Tozzi \& Norman 2001; Rosati, Stefano, \& Norman et al. 2002) 
typically observed for low-temperature X-ray systems. There have been quite a few suggestions that a non-gravitational heating (thus extra
entropy) might prevent the ICM from further collapsing at the cluster core. 
The effect is supposed to be more pronounced
in colder systems whose virial temperature is comparable to the temperature created by this non-gravitational heating,
leading to shallower gas profiles than
those of high-temperature systems (e.g., Balogh et al. 1999; Tozzi \& Norman 2001).
Interestingly, our determination of the surface brightness profile of Lynx-W is much shallower ($\beta=0.42\pm0.07$) than that
of Lynx-E ($\beta=0.71\pm0.12$) (however, Ettori et al. (2004) obtained $\beta=0.97\pm0.43$ for Lynx-W).

The relatively loose distribution of the cluster galaxies in Lynx-W without any apparent BCG defining the cluster center 
leads us to consider another possibility that the system might be dynamically young and the ICM
has not fully thermalized within the potential well. If we imagine that
the ICM is not primordial, but has been ejected from the cluster galaxies at some recent epoch,
it is plausible to expect that the X-ray temperature of the ICM might yet under-represent the depth of the cluster
potential well. Tozzi et al. (2003) investigated the iron abundance in the ICM
at $0.3<z<1.3$ and argued that the result was consistent with no evolution of the mean iron abundance 
out to $z\simeq1.2$. If we assume that, as they suggested, Type Ia SNe are the dominant sources of
this iron enrichment and have already injected their metals into the ICM by $z\sim1.2$,
a significant fraction of clusters at $z\gtrsim 1.2$ may possess dynamically young ICM.

Recently, Nakata et al. (2005) reported with a photometric redshift technique 
the discovery of seven other cluster candidates around
these two Lynx clusters possibly forming a $z\sim1.3$ supercluster. Although
further evidence is needed that the individual clumps are dynamically
bound, the clear enhancement of the red galaxies consistent with the color
at the redshift of the two known Lynx clusters is worthy of our attention.
If they are indeed found to be forming groups/clusters at $z\sim1.3$, but
missed by X-ray observations because of their low X-ray contrast, 
the detailed studies
of these young high-redshift structures will provide a critical benchmark
in testing our understanding of the structure formation as well as the individual galaxy
evolution in the context of different environments.

Deep two band ($i_{775}$ and $z_{850}$ $HST$/ACS imaging of the five out of the seven group/cluster candidates of
Nakata el al. (2005) are scheduled in $HST$ Cycle 14 (Prop. 10574, PI. Mei). Studies
similar to the current investigation will not only test
whether there exist dark matter clumps around the candidate galaxies, but also
quantify the environments for the investigation of the cluster galaxy 
color/morphology evolution.

ACS was developed under NASA contract NAS5-32865, and this research was supported
by NASA grant NAG5-7697. We are grateful for an equipment
grant from Sun Microsystems, Inc. 

Some of the data presented herein were obtained at the W.M. Keck
Observatory, which is operated as a scientific partnership among the
California Institute of Technology, the University of California and
the National Aeronautics and Space Administration. The Observatory was
made possible by the generous financial support of the W.M. Keck
Foundation.  The authors wish to recognize and acknowledge the very
significant cultural role and reverence that the summit of Mauna Kea
has always had within the indigenous Hawaiian community.  We are most
fortunate to have the opportunity to conduct observations from this
mountain.

\clearpage

\end{document}